\documentclass{article}
\usepackage[utf8]{inputenc}
\usepackage[nottoc]{tocbibind}
\usepackage{amsmath}
\usepackage{amssymb}
\usepackage{hyperref}
\usepackage{systeme}
\usepackage{float}
\usepackage{algpseudocode}
\usepackage{algorithm}
\usepackage{mathtools}
\usepackage{adjustbox}
\usepackage{rotating}
\usepackage{amsthm}
\usepackage{caption} 
\usepackage[numbers]{natbib}

\usepackage[left=2.5cm,top=2.5cm,right=2.5cm,bottom=2.5cm]{geometry} 

\title{ZeLiC and ZeChipC: Time Series Interpolation Methods for Lebesgue or Event-based Sampling}
\author{Matthieu Bellucci$^1$, Luis Miralles$^1$, M. Atif Qureshi$^1$, Brian Mac Namee$^1$}
\date{$^1$Centre for Applied Data Analytics Research (CeADAR), \\ University College Dublin, D04 Dublin 4, Ireland.}

\begin{document}

\maketitle

\paragraph{Abstract}

Lebesgue sampling is based on collecting information depending on the values of the signal (e.g. the signal output is sampled when it crosses specific limits). Although the interpolation methods for periodic sampling have been a topic of research for a long time, there is a lack of study in methods capable of taking advantage of the Lebesgue sampling characteristics to reconstruct time series more accurately. Indeed, Lebesgue sampling contains additional information about the shape of the signal in-between two sampled points. Using this information would allow us to generate an interpolated signal closer to the original one. That is to say, the average distance between the interpolated signal and the original signal will be smaller than a signal interpolated with other interpolation methods. In this paper, we propose two novel time series interpolation methods specifically designed for Lebesgue sampling called ZeLiC and ZeChipC. ZeLiC is an algorithm that combines both Zero-order hold interpolation and Linear interpolation to reconstruct time series. ZeChipC is a similar idea, it is a combination of Zero-order hold and PCHIP interpolation. Zero-order hold interpolation is favourable for interpolating abrupt changes while Linear and PCHIP interpolation are more suitable for smooth transitions. In order to apply one method or the other, we have introduced a new concept called tolerated region. ZeLiC and ZeChipC include a new functionality to adapt the reconstructed signal to concave/convex regions. The proposed methods have been compared with the state-of-the-art interpolation methods using Lebesgue sampling and have offered higher average performance. Additionally, we have compared the performance of the methods using both Riemann and Lebesgue sampling using an approximate number of sampled points. The performance of the combination ``Lebesgue sampling with ZeChipC interpolation method" is clearly much better than any other combination.

\section{Introduction}

Nowadays, a lot of time series data is produced, which represents the state of the environment over a period of time \cite{hamilton1994time}. These data points are generally captured by a piece of equipment called a sensor. The sensor can detect different events or changes in the environment and quantify the changes in the form of temperature, pressure, noise, or light intensity, among others. A limitation of collecting data points is the frequency at which the sensor records the changes or events. The more frequently a sensor records a reading, the more expensive the running cost is. Likewise, the less frequent the sensor records the reading, the more difficult it is to capture and reconstruct the original behaviour of the event. 

In practice, all signals have to be sampled because the number of points in a continuous environment is infinite. Sampling is the mechanism that collects the information by setting the frequency of the collected points over a time period. Capturing readings more often is economically more expensive due to the amount of data being stored, transmitted, and processed. The challenge while performing sampling is to preserve the vital information in the less amount of data points so that the objective of recording changes is met.

The periodic or Riemann sampling \cite{hamilton1994time} is a conventional approach of sampling in the time series data. In this approach, the data is captured periodically, i.e. at an equidistant time intervals (such as each second or each microsecond). Even though the approach is simple to implement, the shortcoming is that, when the sampled data fails to indicate changes that happen between the interval (also known as frequency aliasing), sampling needs to be readjusted at a higher frequency, resulting into more data collection. Firstly, making such an adjustment requires manual assessment, and in addition to that, it bears the additional cost concerning more data being generated. Due to this pitfall, many research findings advocated for the use of Lebesgue sampling, instead of Riemann sampling \cite{meng2012optimal}. Furthermore, some authors \cite{aastrom1999comparison} have demonstrated Lebesgue sampling being a more efficient strategy compared to Riemann sampling.

The Lebesgue sampling  \cite{astrom2002comparison}, also known as Event-based sampling, is an alternative sampling strategy to the more popular Riemann sampling strategy. In the Lebesgue sampling, the time-series data is sampled whenever a significant change takes place or when the measurement passes a certain limit \cite{heemels2012introduction}. A few motivating examples of sampling strategy would be: whenever a specific value of the sensor reading crosses a limit, when a data packet arrives at a node on a computer network, or when the system output has changed with a specified amount. 

The overall intuition of the Lebesgue sampling is to save the unnecessary data from being stored, processed, or transmitted which represents either no change or a trivial change compared to the previous data point. The nature of sampling based on events in the Lebesgue sampling is very appealing and natural in many domains where the systems remain constant for an extended period such as wireless communications \cite{imer2005optimal} or systems with an on-off mechanism like those in the satellite control \cite{aastrom1999comparison}.

Increasing the battery life of the sensors and reducing their use \cite{miskowicz2006send}, reducing network traffic by decreasing the amount of information transferred  \cite{zhang2016survey}, or using fewer computer resources while maintaining the same performance \cite{yan2016lebesgue}, are some of the advantages of event-based control over control based in time. By contrast, the management of the systems that implement Lebesgue sampling becomes more complicated \cite{aastrom1999comparison}.

When a time series signal is sampled, generally, the subsequent step is to reconstruct the original signal as accurately as possible \cite{fu2011review}. The interpolation method is one of the well-known criteria to reconstruct the signal by filling the missing values between the range of the discrete set of data points. Despite the significance of the interpolation methods, the challenge of reconstructing the signal remains an important area of research. Moreover, common interpolation methods do not perform well on Lebesgue sampling as demonstrated in this contribution.

In this contribution, it is proposed interpolation methods to reconstruct the time-series data sampled using Lebesgue sampling. To the best of our knowledge, this is the first interpolation method designed exclusively for Lebesgue sampling. The proposed methods have a higher performance because they exploit the particular properties of this kind of sampling.

Two novel interpolation methods are proposed: ZeLiC and ZeChipC. ZeLiC uses \textbf{Ze}ro-order hold and \textbf{Li}near interpolation with specific shape approximation on the basis of \textbf{C}oncavity/\textbf{C}onvexity\footnote{Note that the name ZeLiC is extracted from this combinations of {\textbf{Ze}}ro-order hold, {\textbf{Li}}near interpolation and the new functionality {\textbf{C}}oncavity/Convexity. The same criteria applies to ZeChipC.}. On the other hand, ZeChipC uses \textbf{Ze}ro-order hold and P\textbf{Chip} interpolation with the same \textbf{C}oncavity/\textbf{C}onvexity improvement as ZeLiC.

The rest of the paper is organised as follows. In section 2, it has been elaborated a review of the state-of-the-art interpolation methods and the background on the Lebesgue sampling technique. In section 3, it is presented the proposed interpolation methods: ZeLiC and ZeChipC, along with their simpler versions, ZeLi and ZeChip, which do not observe shape approximation on the basis of Concavity/Convexity. In section 4,  two experiments using 67 different data sets are carried out. The objective of the first experiment is to compare the performance of the proposed methods against that of the state-of-the-art interpolation methods for Lebesgue sampling. The second experiment is used to compare the performance of Lebesgue and Riemann sampling with approximately the same number of samples and using the same methods. This is very useful in order to decide which is the best combination of sampling and interpolation methods when time series need to be sampled. Finally, in section 5, the conclusions of the research are shown along with some possible future directions. Additionally, due to the great extension of the results of the experiments, all the tables are presented in the Appendix section.

\section{State of the art}

In this section, two well-known topics in the domain of time series are described: First, a summary of the state of the art about Event-based or Lebesgue sampling technique. And second, an overview of some popular time series interpolation methods.

\subsection{Lebesgue sampling}

\begin{figure}
\centering
\includegraphics[width=1\linewidth]{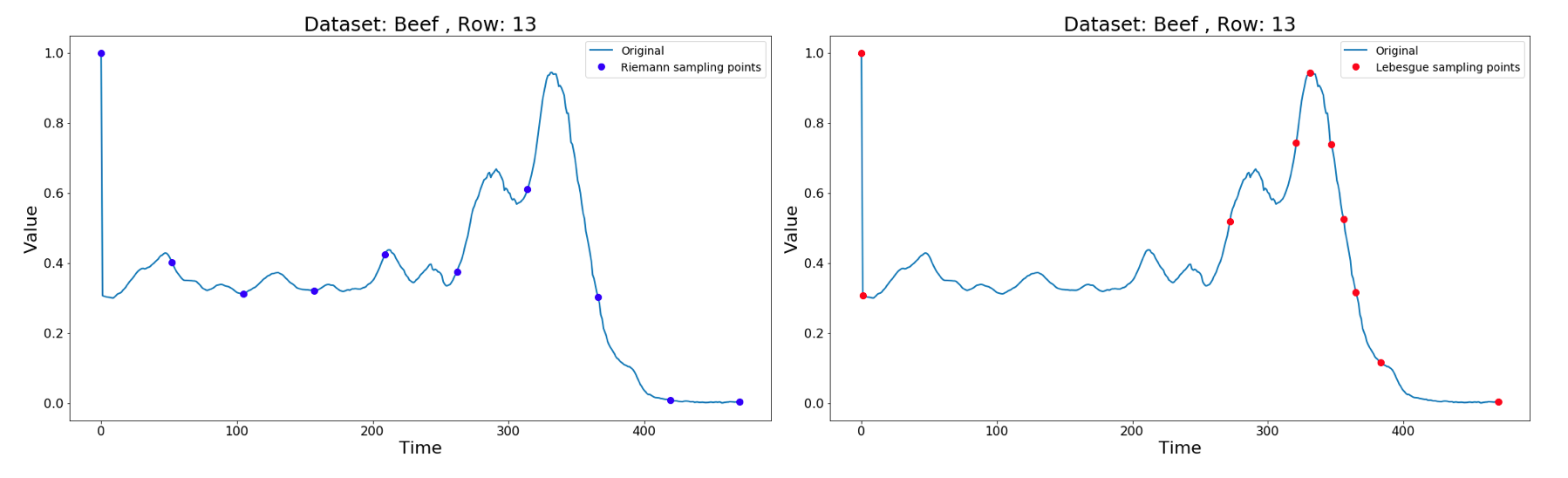}
\caption{Riemann sampling (left) takes points at an equidistant time intervals while Lebesgue sampling (right) does so based on the output value of the signal. In this particular case, when the absolute difference is higher than 0.2.}
\label{fig:Riemann}
\end{figure}

Lebesgue sampling is an alternative to the traditional approach of sampling time series at a constant frequency. Instead of periodically taking samples from a system like in Riemann sampling, the event-based method takes samples only when a predefined event happens as shown in Figure \ref{fig:Riemann}. Some examples of typical events could be a sudden change in the signal, the signal reaching a preset limit, the arrival of a data package, or a change in the state of a system \cite{aastrom1999comparison}. Even though Lebesgue sampling is more accurate than Riemann sampling it is less extended because those systems are more difficult to implement \cite{aastrom1999comparison}.

In recent years, a great interest has aroused in applications implementing event-based sampling. For example, the “Send on Delta” algorithm takes advantage of Lebesgue sampling to reduce the information transmitted by wireless networks in order to increase the lifetime of the sensors batteries. Under this scheme, the sampling is performed only when there is a deviation in the signal higher than the delta value. Results show that using this approach it is possible to increase the lifetime of the sensors without any loss of quality in the resolution of the signals \cite{miskowicz2006send}.

We can find another positive example in the domain of Networked Control Systems (NCSs), where the advantages of Lebesgue sampling become clear. In this type of systems increasing the sampling frequency can be counterproductive since the information load increases and the traffic of the network can collapse the whole system functioning. In the last decade, many NCS have successfully implemented event-triggered control reducing the required resources and the bandwidth of the network \cite{zhang2016survey}.

It is also interesting pointing out the convenience of using event-based sampling in the Fault Diagnosis and Prognosis (FDP). In the last years, it has been increasingly difficult to manage microcontrollers and embedded systems due to the volume of the information collected by sensors and to the complexity of the programs that they implement. Increasing computational resources is not a good solution in the long term since it increases economic costs. Yan et al. \cite{yan2016lebesgue} found an efficient solution to this problem applying the philosophy of ``execution only when necessary" based on Lebesgue sampling, which reduces computational costs substantially without diminishing the performance of the system.

Some research has been done to find the optimal balance between the number of samples and the performance of the system. For example, Andrén, et al. \cite{andren2017event} studied this balance for a linear-quadratic-Gaussian (LQG) control problem setting with output feedback. However, sampling based on changes works well with signals that remain constant for some time and present sudden variations. This is because this kind of sampling captures a higher number of points when the signal has abrupt changes while it does not take points when the signal remains constant. For example, when a natural phenomenon like an earthquake takes place, the sensor values can go from zero to a positive high value in an instant. With sampling based on events, more points of the critical moments would be captured, which gives important information about the behaviour of the phenomenon while if nothing occurs no information will be captured.

Lebesgue sampling can minimise energy consumption, storage space, computation time and the amount of data to be transmitted as it has been claimed in many investigations \cite{wang2014comparison}. It is therefore very interesting for companies who are in charge of monitoring complex systems, as it significantly reduces the expense, without a negative impact on the precision of the measures. 

In summary, the traditional approach for sampling and digital control has been working well in many applications for many years. However, there are new domains where Riemann sampling has major problems that can be easily solved by implementing Lebesgue sampling. That is why in recent years a general interest has been awakened by this new approach \cite{aastrom2003systems}.

\subsection{Time series interpolation methods}

The downsampled time series data can be reconstructed using different interpolation techniques. The interpolation function estimates the missing data points within the range of the discrete set of known data points with the objective of preserving the shape of the original signal before the application of downsampling \cite{lepot2017interpolation}. 

Let $\left(x_i,y_i\right), i=0,\dots,n$ be pairs of real values, where $f$ is the interpolation function, $x_i$ are the indexes of the downsampled data points, and $y_i$ are the values of those points. The objective of the optimal interpolation technique is to satisfy the condition where $f(x_i)$ verifies $y_i$.
 \begin{equation}\label{interp cond}
 f\left(x_i\right)=y_i \text{ for } i=0,\dots,n
 \end{equation}
There are a number of interpolation techniques ranging from simpler ones such as Zero-order hold \cite{de1978practical} and Linear \cite{de1978practical}, to a more complex ones such as Multiquadric \cite{hon1997multiquadric} which is based on radial basis function, Shannon \cite{marks2012introduction} which is based on  Nyquist–Shannon sampling theorem, Lasso \cite{tibshirani2011regression} which is based on regression, Natural Neighbour \cite{boissonnat2002smooth} which is a spatial method and provides smoother approximation compared to simple interpolation technique. Cubic  Hermite spline \cite{keys1981cubic} and Piecewise Cubic Hermite Interpolating Polynomial (PCHIP) \cite{kahaner1989numerical} are interpolation techniques based on splines and cubic function respectively. They are often a preferred choice in the polynomial interpolation.

The relation \ref{interp cond} describes the objective function of an interpolation method. Let $\left(x_i,y_i\right), i=0,\dots,n$ pairs of real values. We want to find a function $f$ (easy to calculate), where $f$ verifies  
 \begin{equation}\label{interp cond}
 f\left(x_i\right)=y_i \text{ for } i=0,\dots,n
 \end{equation}

\subsubsection{Zero-order hold interpolation}

The zero-order hold (ZOH) interpolation is one of the simplest signal reconstruction techniques \cite{de1978practical}. In this technique, the missing values between two sampled points are interpolated with a constant value. This constant value is the same value of the preceding known point before encountering missing values. This technique has several applications in electrical communication and the main advantage is its low computational complexity. However, this interpolation strategy fails to reconstruct continuity or trend in time series with non zero values for the first derivative.

In ZOH, the polynomial $C_i(x)$ is of 0\textit{th} degree. Therefore,  $C_i(x)=c$ where $c$ is constant. Because of \eqref{interp cond}, $C_i\left(x_{i-1}\right)=y_{i-1}$. If in $x_i$, a sudden change is presented, we cannot represent it because $C_{i+1}\left(x_i\right)=y_i\neq y_{i-1}$. Therefore, this interpolation cannot be used neither to represent continuous functions nor to produce a natural curves.

\subsubsection{First order or Linear interpolation}

Linear interpolation is also another popular choice for the reconstruction of a signal due to its simplicity and low computational complexity \cite{de1978practical}. In this method, missing values are reconstructed by fitting a straight line between successive known points. The shortcoming is that the reconstruction of a signal fails to capture any non-linear trend even though the overall known values follow a non-linear trend.

The polynomial $C_i(x)$ is of 1st degree, which means $C_i(x)=ax+b$, and because of \eqref{interp cond}, $C_i\left(x_{i-1}\right)=y_{i-1}$ and $C_i\left(x_i\right)=y_i$, where each two consecutive pair of sampled points are connected using a straight line. The following formulas can be applied to calculate $a$ and $b$, $\forall i=1,\dots,n$
\begin{equation} \label{linear_inter}
     \systeme*{a = \frac{y_i-y_{i-1}}{x_i-x_{i-1}},b=y_i-ax_i}
 \end{equation}
This interpolation gives a continuous but non-differentiable function $f$. Let $C_i(x)=a_ix+b_i$ and $C_{i+1}=a_{i+1}x+b_{i+1}$, we have
 \begin{align*}
 \lim_{x^{-}\to x_i} f'(x) &=  a_i\\
 \lim_{x^{+}\to x_i} f'(x) &= a_{i+1}
 \end{align*}
Therefore, the function is differentiable only if $a_i = a_{i+1},\quad\forall i=1,\dots,n$. We can conclude that for non-linear functions, $f$ is not differentiable with linear interpolation.

Linear interpolation is fast to compute and very intuitive. However, the drawback is the non-derivability of the interpolation at each node, which makes it produce sharp changes in the reconstructed signal.

\subsubsection{Spline interpolation methods}

In the spline interpolation methods, the interpolation function is a particular case of piecewise polynomial. The advantage of this type of method over the first order interpolation is that it reconstructs the signal using non-linear functions (which makes the transitions smoother) and also avoids the Runge phenomenon \cite{atkinson2005theoretical, splineA}. The Runge phenomenon refers to the oscillation at the edges of a given interval while interpolating missing values.

We can define $f$ as $f(x)=C_i(x)$ for $x \in \left[x_{i-1},x_i\right],i=1,\dots,n$ where $C_i$ is a polynomial of small degree. Where $C_i\left(x_i\right)=C_{i+1}\left(x_i\right)=f\left(x_i\right)=y_i$. The most common degrees in terms of interpolation are the first and third degree, which are Linear and Cubic interpolation.

\subsubsection{Third order or Cubic interpolation}

The cubic function is one of the most commonly used spline interpolation methods \cite{keys1981cubic}. It uses a third-degree polynomial in the Hermite form for interpolating missing values as follows $C_i(x)=ax^3 + bx^2 + cx + d$. This interpolation strategy inherits the conditions of Linear interpolation and adds conditions on its first and second derivatives: 
\begin{align*}
C_i\left(x_i\right)&=C_{i+1}=f\left(x_i\right)\\
C_i'\left(x_i\right)&=C_{i+1}'=f'\left(x_i\right)\\
C_i''\left(x_i\right)&=C_{i+1}''=f''\left(x_i\right)
\end{align*}
The strength of this interpolation strategy is the fact that it produces smooth curves in the region of missing values which makes the signal look natural. The drawback of this method is that it can lead to significant errors in the region of reconstruction when there is an abrupt change at the end of an interval. Where the derivative on the nodes (sampled points) must be equal, this change will be presented in the beginning of the next interpolated interval.

\subsubsection{Piecewise Cubic Interpolating Polynomial (PCHIP)}

The PCHIP (Piecewise Cubic Hermite Interpolating Polynomial) interpolation method is based on the same principle as the spline interpolation, but between each point, it fits a cubic polynomial in Hermite's form \cite{kahaner1989numerical}. The sampled points are known as “knots”, PCHIP connects those “knows” independently making a good performance in time and results. The given points have first derivative at the interpolated points, although the second derivative is not guaranteed to be continuous.

\section{ZeLiC and ZeChipC Lebesgue sampling interpolation methods}

In this section, first, we give some information that can be extracted from sampling based on events to develop methods that either need less sampled points to achieve a similar accuracy level than the state of the art methods or given the same amount of points, achieve higher accuracy. Then, we describe in detail the development of ZeLi, the simplest interpolation method proposed for Lebesgue sampling. Following, we describe ZeLiC, which is an improved version of ZeLi with a new functionality to adapt it to convex and concave regions\footnote{The implementation code for the proposed algorithms can be found in:  \url{https://github.com/shamrodia74/ZeLiC}}.  Additionally, mathematical demonstrations to support the assumptions of the developed methods have been included. Finally, we propose a method called ZeChipC which is basically an adaptation of ZeLiC that uses PCHIP instead of Linear interpolation so that it can represent signals with curve regions.

\subsection{Nomenclature}

The purpose of the following list of mathematical nomenclature is to facilitate the readers the understanding of the theory behind the proposed algorithms.

\begin{itemize}
\item $\left(x_i,y_i\right)$: It represents a sampled point, where $x_i$ represents the index and $y_i$ its value.
\item $I_i=\left[x_i,x_{i+1}\right]$: It is the interval that contains all the points between the i\textit{th} sampled point and the next sampled point. This nomenclature is used when continuous interpolation is applied.
\item $x_{i,j}$: It represents the j\textit{th} point to be interpolated in the interval $I_i$. This notation is used to interpolate time series in a discrete manner.
\item \textit{t}: The threshold is used to decide if a given point is captured based on its difference with respect to the last sampled point.
\item $f$:  It is the function that represents the sampled signal, such as $f\left(x_i\right)=y_i$.
\item $p$: It is the function used to interpolate the signal.
\item $R_i = [y_i - t, y_i + t]$: It is the tolerated region a concept described in \ref{sub:introduction}.
\item $\mathbb{P}\left(x \in E\right)$: The probability of the element $x$ to be in the ensemble $E$.
\item Tolerated region = $[(y_i - t), (y_i + t)]$: The tolerated region is defined by the last sampled point. It is calculated as the area covered by two points, which are the value of the last sampled point plus and minus the threshold. If the values of the next points of the signal are inside the tolerated region then the values are not captured.
\item Increased tolerated region = $[(y_i - t)*tolerance\,ratio, (y_i + t)*tolerance\,ratio]$: It is a parameter used to determine if the transition between two points has been smooth or abrupt. For example, ZeLi applies Linear or ZOH interpolation based on this parameter.

\end{itemize}

\subsection{Tolerated region} \label{sub:introduction}

\begin{figure}
\centering
\includegraphics[width=1.0\linewidth]{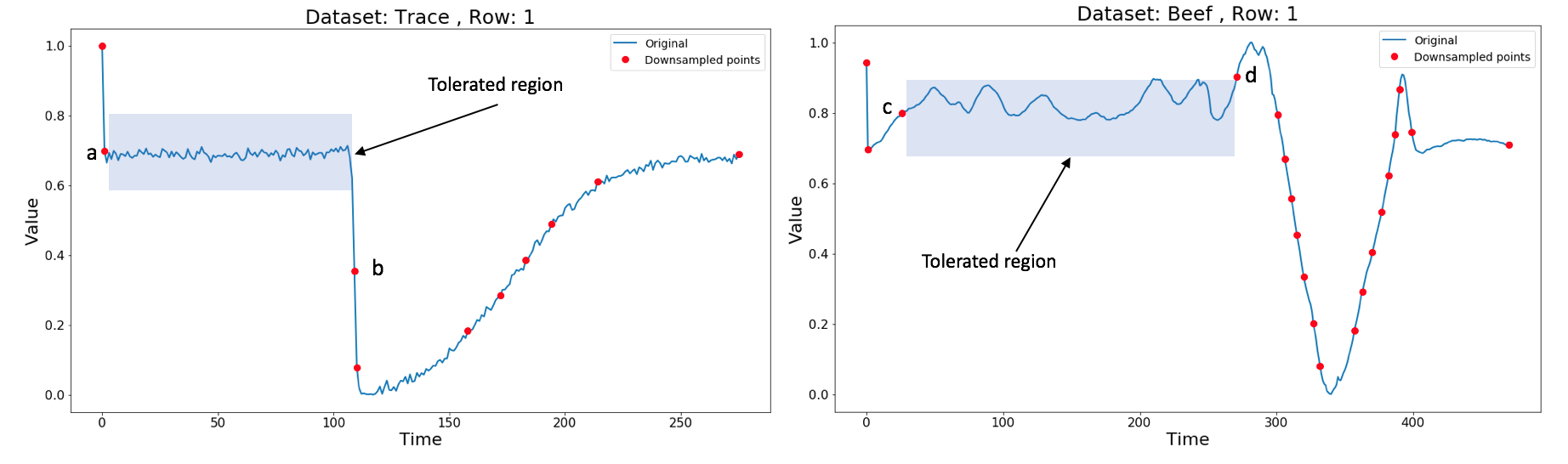}
\caption{The point\textit{ b} is very far away from the tolerated region of the point \textit{a} (Left) , so we can deduce that an abrupt change has taken place. In contrast, the point \textit{d} is very near to the tolerated region of the point \textit{c} (Right), thus it can be inferred that a smooth transition has occurred.}
\label{fig:tolerated_region}
\end{figure}

There are many possible implementations of Lebesgue sampling for time series such as sampling a point when it crosses a preset limit or when the percentage variation is higher than a given threshold.  Our particular implementation of Lebesgue sampling is based on the variation of the signal.  In other words, when the sensor detects that a point has a difference from the previous sampled point higher than a given threshold, then the point is captured. We can express this same idea in mathematical terms in the following way. Let $y_i$ be the first sample of a signal and $t$ the threshold, where $t > 0$. Then, the sensor captures the next point called $y_{i+1}$ if and only if:
\begin{equation}\label{percentage rule}
    \left|y_{i+1}-y_i\right|\geq t
\end{equation}
Lebesgue sampling indirectly gives information on the behaviour of the signal between the two consecutive samples. It can be deduced that all the points between two pair of consecutive sampled points (not necessarily consecutive in time intervals) are inside an interval delimited by the threshold as shown in Figure \ref{fig:tolerated_region}. We know this because if the value of a point of the interval was out of the interval this point would have been captured. Based on that simple deduction a new concept called tolerated region because points are not sampled while the signal stays inside this allowed region. The tolerated region can be defined as $[y_i - t, y_i + t]$, where $t$ (threshold) is the maximum allowed value of change. 

As shown in Figure \ref{fig:tolerated_region}, if the difference between two consecutive sampled points \textit{a} and \textit{b} (left) of a time series is very large in a given interval, we know that an abrupt change has happened (otherwise the point would have been collected earlier). On the other hand, if the difference between the sampled points is very small, like in the case of \textit{c} and \textit{d} (right), it is quite probable that a smooth change has taken place. This simple principle is the basis on which the interpolation algorithms for Lebesgue sampling have been developed.

We can express this same idea in mathematical terms in the following way. Let $x_i,\,i=0,\dots,n$ the position of the sampling points, let $y_i=f\left(x_i\right)$ be the position of the signal at each sampling point, let $(x,y)$ be a point between two sampled points, that is to say $x \in \left[x_i,x_{i+1}\right]$, and let $R_i$ be the tolerated region where $y$ can be: $R_i=\left[y_i-t,y_i+t\right]$. Then, with a periodic sampling, we would have that the probability of all the points being in the tolerated region is less than one, $\mathbb{P}\left(y \in R_i\right)\leq 1$. Whereas with Lebesgue sampling all the points are inside that region, $\mathbb{P}\left(y \in R_i\right)=1$. In other words, any $y$ in the interval $I$ between the two points is inside the tolerated region. We can therefore significantly reduced the region of the possible values of $y$ when performing the interpolation and therefore reducing the error when comparing the original signal with the reconstructed one.

\subsection{ZeLi interpolation algorithm}\label{sec:zeli-interpolation}

From the information that can be extracted from the tolerance region, we will develop a set of methods to interpolate time series sampled with the Lebesgue approach. The simplest method and the first that is going to be explained is called ZeLi. The rest of the methods are improvements with respect to this first method. Along with the explanation of the methods, it is presented the mathematical demonstrations to proposed methods more rigorous.

\subsubsection{Combination of Zero-order hold and Linear interpolation} \label{Zeli description}

\begin{figure}
\centering
\includegraphics[width=1.0\linewidth]{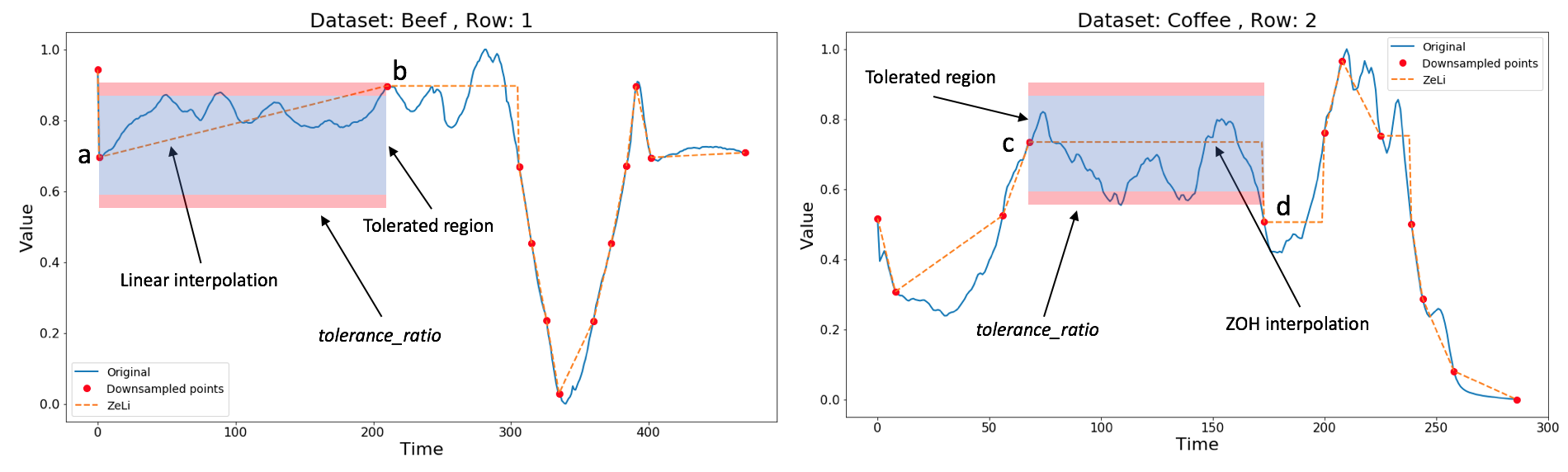}
\caption{ZeLi combines ZOH with Lineal interpolation. Between \textbf{a} and \textbf{b} ZeLi applies Line interpolation (left) while between the points \textbf{c} and \textbf{d} it applies ZOH interpolation (right).}
\label{fig:zeli_interpolation}
\end{figure}

ZeLi interpolation combines \textbf{Ze}ro-order hold interpolation and \textbf{Li}near interpolation, which explains the origin of its name, to reconstruct the original signal from the sampled signal as shown in Figure \ref{fig:zeli_interpolation}. To decide whether to apply ZOH or Linear interpolation, the \textit{tolerance ratio} parameter is used. The \textit{tolerance ratio} is a constant value higher than multiplies the interval of the tolerated region, that is: $[(y_i - t)*tolerance\,ratio, (y_i + t)*tolerance\,ratio]$, creating a new interval called increased tolerated region.

Therefore, the ZeLi algorithm contemplates two possible cases
\begin{enumerate}
\item  If the examined point is outside of the increased tolerance region ( the tolerance region multiplied by the {tolerance ratio}), then ZOH interpolation is used.
\item If the examined point is inside of the increased tolerated region, then Linear interpolation is used.
\end{enumerate}

The justification of the algorithm is as follows. We know that all the points between two captured points should be in the tolerated region. if the difference between the values of the points is small, it is quite possible that between the two sampled points the signal follows a linear trend with small variations around it, therefore Linear interpolation is used. On the other hand, to minimise the error when the difference between the two sample points is great (an abrupt change is presented), we interpolate all the points using ZOH. Although ZOH interpolation does not represent continuous signals in a smooth way, it is a very effective mechanism since it satisfies that all the values of the interval between a given point and the next point [a,b) are in the tolerated region. ZOH interpolation minimises the error because it keeps all the values in the middle of the tolerated region $R_i$, dividing by two the maximal possible error. 

A visual intuition of this method can be seen in Figure \ref{fig:zeli_interpolation}. When the signal crosses the threshold in a continuous way (left) Linear interpolation would be used. This is because it can be assumed that the values of the previous points close to \textit{b} will have similar values (obviously, the further those points are from \textit{b}, the less probable this assumption would be). By contrast, if an abrupt change is presented  (right), then the last but one point and all the previous points between \textit{c} and \textit{d} were somewhere in the limited region $R_i$, but we cannot deduce a trend, because the change is so abrupt. Because of this lack of information, we choose to minimise the maximal error by using ZOH interpolation. Since ZOH interpolation uses a constant line (same values for y-axis for all the points of the region) followed by a straight line to interpolate.

\subsubsection{Mathematical definition for ZeLi interpolation method}

We can express this same idea in the following way. Let $f$ be a function  representing a time series. Let $\left(x_i,y_i\right),\,i=0,\dots,n$ be a sampled point, where $y_i=f\left(x_i\right)$, sampled with a threshold sampling. We want to interpolate the time series using the extracted information from the sampling. Let $p$ be the polynomial that interpolates the signal on the interval $I_i=\left[x_i,x_{i+1}\right]$, such as $p\left(x_j\right)=f\left(x_j\right)$ $\forall j=i,i+1$. Let $R_i=\left[y_i-t,y_i+t\right]$ the tolerated region such as $y=f(x) \in R_i \: \forall x \in I_i$. 

If $p$ is of 1-order or higher, it cannot be guaranteed that $f(x) \in R_i \: \forall x \in I_i$. Thus, ZOH interpolation is the only spline interpolation which satisfies the condition that $y$ remains in $R_i$. 

From the study of the variation between $y_i$ and $y_{i+1}$, we can distinguish two cases:
\begin{enumerate}
\item The difference between two consecutive sampled points is slightly higher than the threshold: $\left|{y_{i+1}-y_i}\right| \gtrapprox t $.
\item The difference between two consecutive sampled points is larger than the threshold: $\left|y_{i+1}-y_i\right| >  t$.
\end{enumerate}

In the first case, we know that if Linear interpolation is applied in the interval $I_i$ of the signal, that is to say $p(x)=ax+b$, then most of the points of the interval will be included in the tolerated region $R_i$, $\forall x \in I_i, \, \left|p(x)-y_i\right|<t$.

In the second case, we cannot guarantee that if we use Linear interpolation, the interpolated points will be contained in the tolerated region, that is to say, we can not guarantee that $\forall x \in I_i, f(x) \in R_i$. Moreover, if the difference is much greater than $t$, we know that all the previous points of the signal being sampled were in the permitted interval. The abrupt change happened at the point $x_{i+1}$ and not before. We also know that the signal is not continuous at this point, because of the sudden change at $x_{i+1}$. Therefore, it is required to use an interpolation method able to represent a non-continuous function,  that is why we use ZOH.

Let us formalise the condition to apply ZOH or Linear interpolation. Although the condition is defined for our Lebesgue implementation, it can be easily extended to other implementations of event-based sampling.

As it has been explained in \ref{sec:zeli-interpolation}, we will use ZOH when the difference is greater than the threshold $t$.
\begin{align*}
    &\left|p(x)-y_i\right| > t
    \implies \left|p(x)-p\left(x_{i}\right)\right| > t
    \implies \left|ax + b - ax_i - b\right| > t
    \implies  \left|a\left(x-x_i\right)\right| > t\\
    \implies &\left|a\right| \left|x-x_i\right| > t
    \implies  x-x_i > \frac{t}{\left|a\right|} \quad \forall x \in I_i, x \geq x_i
    \implies x > \frac{t}{\left|a\right|} + x_i
\end{align*}

The algorithm \ref{alg:zeli} for ZeLi interpolation on $I_i$ is rather simple. Let us suppose that we are analysing signals in the discrete-time domain, let $x_{i,j}$ for $j=0,\dots,n$ be the points in $I_i=\left\{x_i,x_{i,0},\dots,x_{i,n},x_{i+1}\right\}$. The condition to apply ZOH interpolation means that for any point of interpolation the following point will have a higher value on the x-axis. Thus, $\forall x \in I_i-\left\{x_{i+1}\right\},\, x_{i,n}\geq x $. That is why it is only required to verify that
\begin{equation}\label{limit_cond}
x_{i,n} > \frac{t}{\left|a\right|} + x_i
\end{equation}

\paragraph{PROOF}

In the following lines, we demonstrate that at least one point of the interpolation is outside the tolerated region $R_i$ by checking an inequality on a single point. This way, we can avoid checking over all the interpolated points which makes the algorithm faster. We assume that $p(x)-y_i>0$ for convenience. If it were negative, the result would be equivalent. The condition for $p(x)$ to be in the tolerated region is $p(x) \leq y_i + t$, which is equivalent to $ p(x)-y_i\leq t$.

\begin{align*}
    p(x)-y_i&\leq t \\
     ax - ax_i &\leq t \quad \text{because}\: y_i = p \left(x_i\right)\\
     x &\leq \frac{t}{a} + x_i
\end{align*}
Because we work in a discrete space, we have $I_i = \{x_i, x_{i,0},x_{i,1},\dots , x_{i,n}\}$ and therefore $\forall x \in I_i$, $x\leq x_{i,n}$. Thus, to check if all points are in the tolerated region, we simply have to check that 
\[
x_{i,n} < \frac{t}{a} + x_i
\]
which concludes the proof.
\qed

\begin{algorithm}
\caption{ZeLi algorithm}\label{alg:zeli}
\textbf{Input: }
$signal\_downsampled, threshold, tolerance\_ratio$
\begin{algorithmic}[1]
\State tolerance = threshold * tolerance\_ratio
\State $signal\_reconstructed$ = List()
\For{\texttt{each pair of points (a,b) in $signal\_downsampled$}}
    \If{$abs(a.value - b.value) < tolerance$}
        \State $upsampled = linear\_interpolate(a,b)$
    \Else
        \State $upsampled = zero\_interpolate(a,b)$
    \EndIf
    \State $signal\_reconstructed$ = $signal\_reconstructed + upsampled$
\EndFor
\State \Return $signal\_reconstructed$
\end{algorithmic}
\end{algorithm}

\subsection{ZeLiC interpolation method}

One of the disadvantages of using simple methods such as linear interpolation and ZOH is that they are not able to adapt properly to the signal when there is a change in the sign of the slope. Therefore, we have developed a new functionality with the aim of improving the performance of ZeLi when convex/concave regions are presented. Next, the mathematical development and the implementation of this improvement are presented.

\subsubsection{Convex/Concave regions on time series}

The shape of the signal when a slope change is presented is generally a convex or concave region. When Lebesgue sampling is used, a lack of information about the shape of the signal is presented when the signal changes the sign of the slope between two sampled points $x_i$ and $x_{i+1}$, and that change takes place inside the tolerance region (i.e. before the signal hits the threshold). Additionally, neither Linear nor ZOH are able to adapt to convex or concave regions of a signal since those two methods connect the pairs of points individually. 

In order to improve the adaptation of the proposed methods, we have studied the case of convexity and concavity; inflexion points of a function. The same conclusions apply to concavity, but the formulas are slightly adapted (with opposing signs). Expressed in mathematical terms could be: 

A function is convex if the line segment between any two points on the graph of the function lies above or on the graph. We can express this same idea in mathematical terms in the following way. A signal $f$ is convex on $I_i$ if and only if
\begin{equation}\label{convex_der}
    \forall x \in I_i, \, f''(x)>0
\end{equation}
or
\begin{equation} \label{convex_seg}
    \exists \lambda \in \left[0,1\right], \quad f\left( \lambda x_i + \left(1-\lambda\right)x_{i+1} \right) \leq \lambda f\left(x_i\right) + \left(1-\lambda\right)f\left(x_{i+1}\right)
\end{equation}

Let $y_{i-1}\geq y_i$ and $y_i \leq y_{i+1}$, then, the global slope of $f$ on $I_{i-1}$ is negative, and the global slope on $I_i$ is positive. From that information the following table can be deduced:

\begin{table}[H]
\begin{center}
\begin{tabular}{|l|lll|}
\hline
$t$ & $x_i$                         & $x$                         & $x_{i+1}$    \\ \hline
$f\left(t\right)$            & \multicolumn{1}{l|}{$\searrow$} & \multicolumn{1}{l|}{}    & $\nearrow$ \\ \hline
$f'\left(t\right)$     & \multicolumn{1}{l|}{$-$}        & \multicolumn{1}{l|}{$0$} & $+$        \\ \hline
\end{tabular}
\end{center}
\end{table}

We can see that the derivative is globally increasing on $I_i$, thus we have this new table:

\begin{table}[H]
\begin{center}
\begin{tabular}{|l|ll|l|}
\hline
$t$   & $x_i$                         & \multicolumn{2}{l|}{$x_{i+1}$}    \\ \hline
$f\left(x\right)$              & \multicolumn{1}{l|}{$\searrow$} & \multicolumn{2}{l|}{$\nearrow$} \\ \hline
$f' \left(t\right)$       & \multicolumn{1}{l|}{$-$}        & \multicolumn{2}{l|}{$+$}        \\ \hline
$f' \left(t\right)$       & \multicolumn{1}{l|}{$\nearrow$} & \multicolumn{2}{l|}{$\nearrow$} \\ \hline
$f'' \left(t\right)$ & \multicolumn{1}{l|}{$+$}        & \multicolumn{2}{l|}{$+$}        \\ \hline
\end{tabular}
\end{center}
\end{table}

If we assume that the function $f$ follows the trend of the sampled points, that is to say, if the values of $f$ for a particular region are decreasing and then, at some point, they start increasing, we can assume that $f$ is convex on the interval $I_i$. At this point, it is important to remind that this is based on assumptions, we do not have enough information to support the assumptions, except the values and position of the sampled points. 

Let us see what would make the assumption wrong. If the signal is convex, we have \eqref{convex_seg} which means that the true values of the signal are all under the Linear interpolation, that is to say, $\forall x \in I_i$
\begin{equation*}
    f(x) \leq p(x)
\end{equation*}
where $p$ is the linear interpolation on $I_i$.

\paragraph{PROOF}

We want to demonstrate that if a function $f$ is convex, then all the true values of the signal, represented by a function $f$ are under the Linear interpolation, represented by $p$, that is to say $f(x) \leq p(x)$.

Let $p(x)=ax+b$ such as $p(x_i)=f(x_i)=y_i$ and $p(x_{i+1})=f(x_{i+1})=y_{i+1}$. Let $\lambda \in \left[0,1\right]$ and $x=\lambda x_i + \left(1-\lambda \right) x_{i+1} \in I_i$. We have $f(x)=f\left(\lambda x_i + \left(1-\lambda \right) x_{i+1}\right)$
\begin{align*}
p(x) &= p\left(\lambda x_i + \left(1-\lambda \right) x_{i+1}\right)
     = a\left(\lambda x_i + \left(1-\lambda \right) x_{i+1}\right) +b\\
     &= \lambda \left(ax_i + b \right) + \left(1-\lambda \right)\left(ax_{i+1} + b\right)
     = \lambda p\left(x_i\right) + \left(1-\lambda \right)p\left(x_{i+1}\right)\\
     &= \lambda f\left(x_i\right) + \left(1-\lambda \right)f\left(x_{i+1}\right)\\
\end{align*}
Using \eqref{convex_seg}, we conclude that if $f$ is convex, then $f(x) \leq p(x)\: \forall x \in I_i$
Therefore, if the signal is not convex, that means that 
\[
\exists x \in I_i, f(x) > p(x)
\]
\qed

Now, let $t$ be the threshold chosen for the sampling, we then know that: 
\[
\forall x \in I, f(x) \in \left[y_i-t,y_i+t\right]=R_i
\]
If a signal is not convex, that means that 
\begin{equation} \label{not convex}
\exists x \in I_i, p(x)<f(x)<f\left(x_i\right)+t
\end{equation}

Let us calculate the length $d(x)$ of the interval $\left[p(x),y_i+t\right]$ for $x \in I_i$. This will give us an idea of the probability of making a false assumption, that is to say, assuming the signal is convex when it actually is not.

\begin{align*}
    d(x)    &= y_i+t - p(x) = p\left(x_i\right) + t - p(x)\\
            &= ax_i + b + t - ax - b\\
            &= -ax + \underbrace{ax_i + t}_{c}
            = -ax + c 
\end{align*}

According to that , $d(x)$ is a polynomial of 1st degree, where $a>0$ because of the fact that $y_i < y_{i+1}$ and $x_i < x_{i+1}$. Therefore, $d(x)$ is decreasing.

Let $R_i$ be the area to which the points may belong, $R_i$ is a rectangle of length $x_{i+1}-x_i$ and of height $2tf\left(x_i\right)$.
We have that $\forall x \in I_i,\mathbb{P}\left(f(x) \in R_i\right)=1$ and we want to know the probability that $f(x)$ is not convex, that is to say, the probability that $\forall x \in I_i, f(x) \in D$, where $D$ is the region defined by $\forall x \in I_i,\left[p(x),y_i+t\right]$.

The area of $D$ is defined as:
\begin{align*}
D_{area}    &= \int_{x_i}^{x_{i+1}} d(x)dx\\
            &= \frac{d\left(x_i\right)\left(x_{i+1} - x_i\right)}{2}
\end{align*}
and the area of $R_i$ is:
\begin{align*}
    R_{i\: area} &= \left(y_i+t - \left(y_i-t\right)\right)\left(x_{i+1} - x_i\right)= 2t\left(x_{i+1} - x_i\right)\\
    d\left(x_i\right) &= y_i+t - y_i= t\\
    \implies R_{i\: area} &= 2d\left(x_i\right)\left(x_{i+1}-x_i\right)= 4D_{area}
\end{align*}

Therefore $\forall x \in I_i, \mathbb{P}\left(f(x) \in D \right) = 0.25$, which means that we have a 25\% chance of making a false convexity assumption and a 75\% of making a right assumption.

The shape of the area where the assumption can be wrong is a right triangle, with its hypotenuse being the Linear interpolation. Because the function $d$, that represents the length of the interval where we can make a wrong assumption is decreasing, we conclude that the assumption is more likely to be wrong at the beginning of $I_i$ than in the end.

\subsubsection{Adding a convexity/concavity assumption to ZeLi} \label{ssec: convexity}

Let us assume that the signal $f$ is convex on the interval $I_i$, $y_{i-1} \geq y_i$ and $y_i \leq y_{i+1}$ . This greatly reduces the possible position of the points on the interval $I_i$. With this hypothesis, we deduce that there exists a point $x$ in $I_i$ where the derivative of $f$ is zero. Indeed, its derivative is increasing thanks to the convexity assumption, and the derivative is negative before $I_i$ (because we have $y_{i-1} \geq y_i$) and positive after $I_i$ (because $y_i \leq y_{i+1}$). Therefore, we could interpolate with one line going from the first point to this point where the derivative is zero, and then another line from this point to the last. 

In order to minimise the maximal error for interpolating convex regions, we choose this new point to be in the middle of $I_i$ and calculate it with the following formula: 
\begin{equation} \label{new point}
x_{i+\frac{1}{2}} = \frac{x_i+x_{i+1}}{2}
\end{equation}

and we choose its value in the y-axis to be the middle of the interval where the points can be, that is to say, in-between the Linear interpolation, and the lower bound of $R_i$ : $y_i-t$. Therefore, we have
\begin{equation} \label{convex point}
    y_{i+\frac{1}{2}}=\frac{p\left(x_{i+\frac{1}{2}}\right)+y_i-t}{2}
\end{equation}

Then, we apply Linear interpolation to interpolate the signal in the convex region. Let $p_i(x)$ be the interpolated signal on $I_i$, we set 
\[
p_i(x) =
\begin{cases*}
    p_{i^-}(x) & if $x < x_{i+\frac{1}{2}}$\\
    p_{i^+}(x) & if $x \geq x_{i+\frac{1}{2}}$
\end{cases*}
\]
This solution has a limitation when ZOH needs to be applied (see Case II described in \ref{Zeli description}). In fact, when ZOH was applied, that meant that a peak was detected. But with the algorithm we use with the convexity assumption, this peak is not necessarily represented, therefore we need to add another step to ensure we have this peak. When an abrupt change takes place it is not recommendable using Linear interpolation because it will interpolate the signal outside of the tolerated region $R_i$. 

This limitation can be solve by adding another interpolation point, $x_{i,n}$, before the last point $x_{i+1}$. If it weren't for the convex assumption, we would have used ZOH interpolation. Therefore, we choose $y_{i,n}=y_i$, because it is the value it would have had without the convex assumption.

\begin{figure}
\centering
\includegraphics[width=0.7\linewidth]{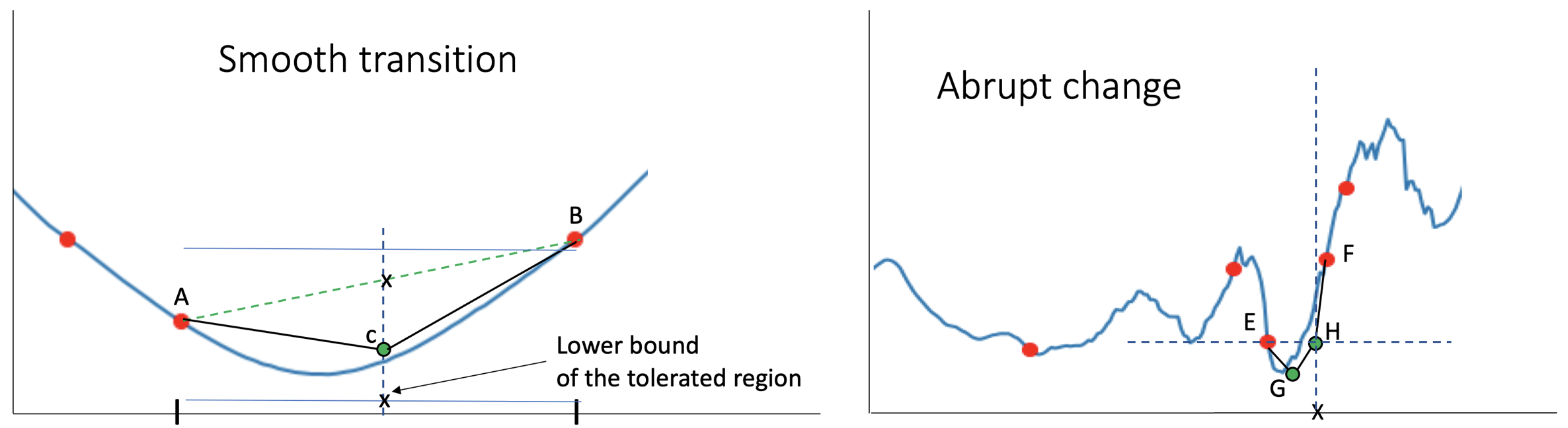}
\caption{When there is a soft transition (left) only one new point is calculated. When an abrupt change occurs (right) we calculate another point for a better adaptation.}
\label{fig:concavity-convexity}
\end{figure}

We call $p_{i^-}(x)$ the Linear interpolation between the points $\left(x_i,y_i\right)$ and $\left(x_{i+\frac{1}{2}},y_{i+\frac{1}{2}}\right)$ and $p_{i^+}(x)$ the Linear interpolation between the points $\left(x_{i+\frac{1}{2}},y_{i+\frac{1}{2}}\right)$ and $\left(x_{i,n},y_{i,n}\right)$. We can write the interpolation function $p_i(x)$ as 
\[
p_i(x) =
\begin{cases*}
    p_{i^-}(x) & if $x_i \leq x < x_{i+\frac{1}{2}}$\\
    p_{i^+}(x) & if $x_{i+\frac{1}{2}} \leq x < x_{i+1}$\\
    y_{i+1} & if $x = x_{i+1}$\\
\end{cases*}
\]
We can see graphically, which is more intuitive, this same methodology in Figure \ref{fig:concavity-convexity}. In the first case (left) the first point for the doing interpolation A = ($A_x$, $A_y$), where $A_x$ is the value in the x-axis and $A_y$ is its value in the y-axis. Let's call the point on the right B = ($B_x$, $B_y$). 

To calculate the new point C, we have to calculate its value on the x-axis and its value on the y-axis:
\begin{itemize}
\item $C_x$: The value is the middle between $A_x$ and $B_x$, that is to say ($A_x$+$B_x$)/2.
\item $C_y$: It is the middle between the points of the lower bound of the tolerated region in $C_x$ (because here the signal is convex) and the value of the linear interpolation (the line that connects the two points) in $C_x$. 
\end{itemize}

In the second case (right), the first new point G is calculated in the same way as C was calculated in the other example. The second new point H is calculated as follows. The x-value, $H_x$, is the point just before $F_x$ in the discrete domain, this is: $H_x$ = $F_x$ – 1. And the value of the y-axis of the point H, $H_y$ is the same as that of the point E, this is, $H_y$ = $E_y$.

This solution adapts the interpolation to the convexity of the function and at the same time, to the abrupt change happening at $x_{i+1}$. The same rules of the concavity assumption can be applied to the convexity assumption with the exception that the sign of the shape is just the opposite. Therefore, adapting the formulas is very easy:

\eqref{convex_seg} becomes
\begin{equation} \label{concave_seg}
        \exists \lambda \in \left[0,1\right], \quad f\left( \lambda x_i + \left(1-\lambda\right)x_{i+1} \right) \geq \lambda f\left(x_i\right) + \left(1-\lambda\right)f\left(x_{i+1}\right)
\end{equation}
We have the same properties and the same implementation that in the Convexity regions. The only difference is the value of $y_{i+\frac{1}{2}}$, the value of the point we added to follow the convexity condition. We now have to calculate it so the signal is concave. Let $l$ be the Linear equation on $I_i$ the new value for $y_{i+\frac{1}{2}}$ is:
\begin{equation} \label{convace_point}
    y_{i+\frac{1}{2}} = \frac{l\left(x_{i+\frac{1}{2}}\right) + y_i+t}{2}
\end{equation}

\begin{figure}
\centering
\includegraphics[width=0.6\linewidth]{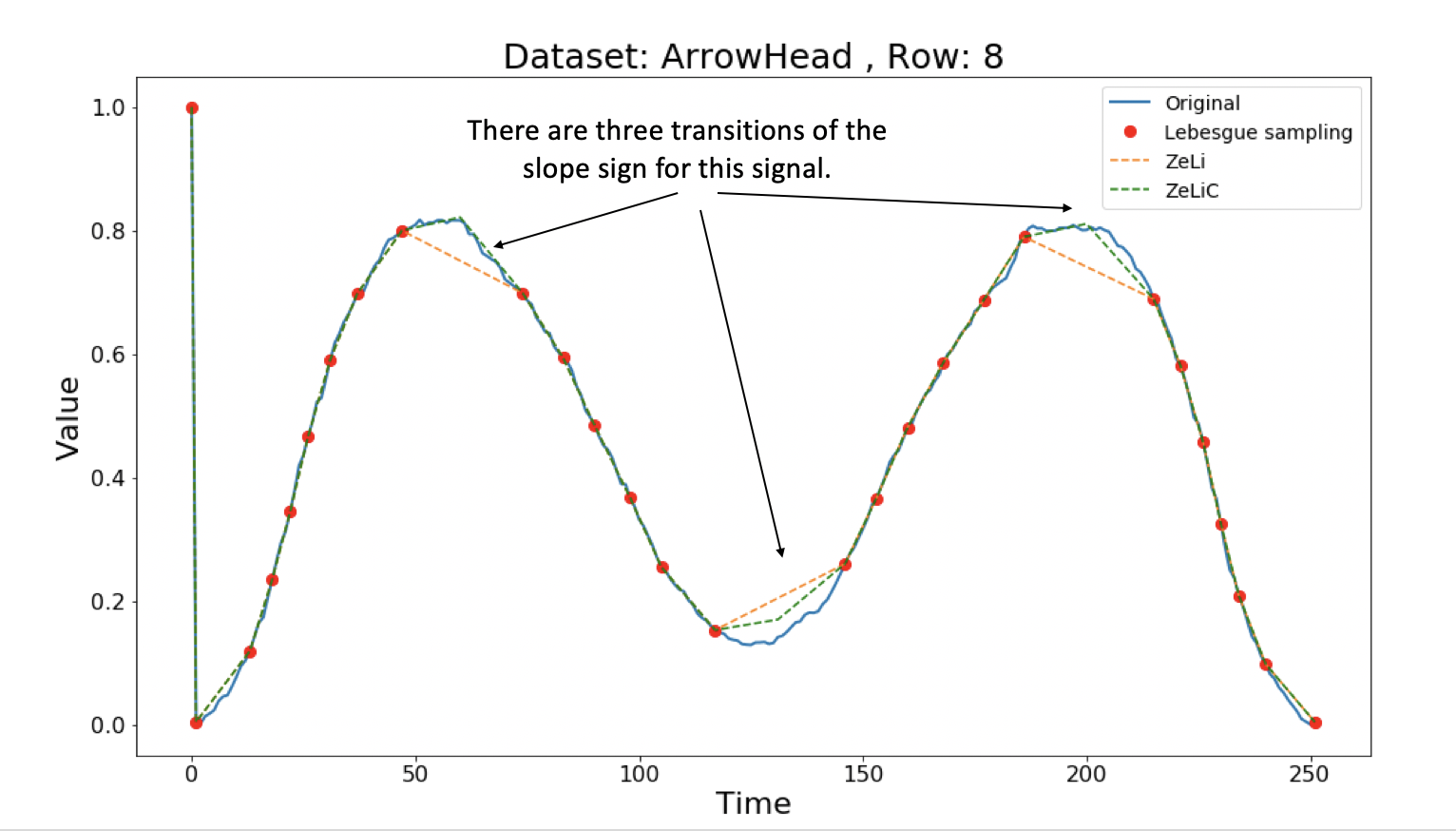}
\caption{ZeLiC is able to follow the shape of the signal with respect to ZeLi thanks to the convexity/concavity condition using a threshold of 0.10}
\label{fig:Concavity}
\end{figure}

\subsubsection{Parameters of the convexity/concavity functionality} \label{convexity}

To reduce the probability of making a false assumption, we added three conditions that have to be met to make the assumption of convexity or concavity. The goal of adding new conditions is to exclude those cases when the assumption is less likely to be true. The downside of doing this is that we will have more false negatives; those cases when is convexity/concavity is not applied and actually, it should be applied. 

To handle the new restrictions for doing a convexity assumption, we have defined three parameters regarding the distances between the points $x_{i-1}$, $x_i$ and $x_{i+1}$. As discussed. The three parameters included in the algorithm \ref{alg:zelic} are the followings.

\begin{itemize}
\item \textit{previous\_distance}: It refers to the minimum distance between $x_{i-1}$ and $x_i$. We want the distance $d=x_i-x_{i-1}$ to be inferior to a chosen threshold. If $d$ is very small, that means that on a small region, the sensor got triggered twice. This happens in signals with very frequent variations. In those cases, it is better not to assume convexity because the signal is highly unpredictable.
\item \textit{subsequent\_min\_distance}: It refers to the minimum distance between $x_i$ and $x_{i+1}$. It has the same purpose as the \textit{previous\_distance} condition. If it is too small, then the interpolation would not benefit from the convexity assumption. It might even cause a bigger error than if we used Linear interpolation or ZOH.
\end{itemize}

The best value for a parameter depends on the shape of the time series for each data set. Therefore, each dataset needs to be optimised individually. As a general idea, both parameters \textit{previous\_distance} and \textit{subsequent\_min\_distance} should have similar values to be consistent. 

That is to say, if the previous distance parameter, $x_i - x_{i-1}$ should not be smaller than a chosen value, called $l$, then it means that this $l$ is the minimum value where we consider the signal to be predictable and therefore assume convexity.

For example, if we set \textit{subsequent\_min\_distance} with a value far larger or smaller than $l$, this would mean that we have significantly changed the limit where we consider the signal predictable. This means that we would assume convexity on $I_{i-1}$ but then, on $I_i$, we change our mind and consider the same distance to be too small to assume convexity. In short, these parameters have to be set by the user, depending on the shape of the signals sampled.

\begin{algorithm}
\caption{ZeLiC algorithm}\label{alg:zelic}
\textbf{Input: }
$signal\_downsampled, threshold, tolerance\_ratio$
\begin{algorithmic}[2]
\State tolerance = threshold * tolerance\_ratio
\State $signal\_reconstructed$ = List()
\For{\texttt{each pair of consecutive points (a,b) in $signal\_downsampled$}}
    \State $is\_convex = change\_ in\_ slope(a,b)$ AND $(b.idx - a.idx > next\_dist)$ \\ AND $(a.idx - (a-1).idx > previous\_dist)$
    \If{NOT $is\_ convex$}
        \If{$abs(a.val - b.val) > tolerance$}
            \State $upsampled = linear\_interpolate(a,b)$
        \Else
            \State $upsampled = zero\_interpolate(a,b)$
        \EndIf
    \Else
        \State $aux_1 = tolerated\:region\:lower\:bound\:value $
        \State $aux_2 = value\:of\:the\:line\:between\:a\:and\:b\:in\:c.idx$
        \State $c.idx = (a.idx - b.idx)/2$
        \State $c.val = (aux_1 + aux_2)/2 $
        \If{$abs(a.val - b.val) < tolerance$}
            \State $upsampled = linear\_interpolate(a,c,b)$
        \Else
            \State $d.val = a.val$
            \State $d.idx = b.idx - 1$
            \State $upsampled = linear\_interpolate(a,c,d,b)$
        \EndIf
    \EndIf
    
    \State $signal\_reconstructed$ = $signal\_reconstructed + upsampled$
\EndFor
\State \Return $signal\_reconstructed$
\end{algorithmic}
\end{algorithm}

\subsection{ZeChip interpolation algorithm}

\begin{figure}
\centering
\includegraphics[width=1.0\linewidth]{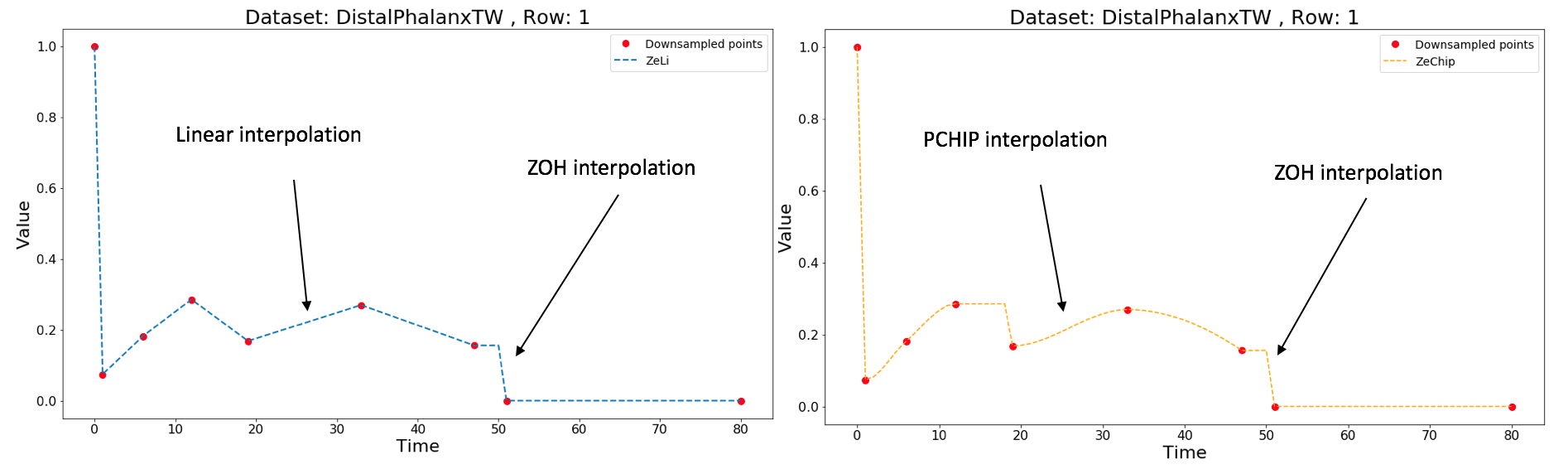}
\caption{(Left) ZeLiC is not able to represent curve lines since it is based on linear interpolation, nevertheless, ZeChip (Right) is able to represent curve lines.}
\label{fig:ZeLiC_limitations}
\end{figure}

ZeLi is based on a combination of ZOH and Linear interpolation and in consequence, it shares some limitations of that this type of interpolation. ZeLi is able to approximate time series with a high precision when they are composed of straight lines, that is to say, signals for whose first and second derivative present very constant values. However, when the signal function has curved lines, it is not possible to represent it using ZeLi. In Figure \ref{fig:ZeLiC_limitations} (left) we can see that ZeLiC is not able to follow the curvature of the line. 

As it can also be observed in \ref{fig:ZeLiC_limitations} (right), we can apply the same idea of combining two interpolation methods as in ZeLi but replacing Linear interpolation with PCHIP interpolation. This new method is called ZeChip and is able to adapt much better to those signals that present curves regions. In addition, the new method will include the advantages of PCHIP; a fast and powerful interpolation method allows to represent regions using curved lines and obtaining a great precision. One of the shortcomings of ZeChip with respect to ZeLi is that ZeChip, due to the fact that uses PCHIP interpolation instead of Linear interpolation, has a higher complexity cost than ZeLi, and therefore, for the same given points, ZeChip will take more time to generate the interpolated signal than ZeLi.

\section{Experiments}

So far, we have discussed our proposed methods from a theoretical perspective but to have certain evidence that our contribution can have a meaningful impact, we need to demonstrate that the proposed methods have a better general performance than that of the state-of-the-art. To this end, we have decided to perform the experiments using a large number of databases. We want to test the performance of our models against other interpolation methods under Lebesgue sampling. Besides that, we want to compare the performance of our method with Lebesgue sampling with the performance of other interpolating methods with Riemann sampling with a similar number of samples, so we can recommend that approach when time series need to be sampled.

\subsection{Preparation of the experiments}

\begin{figure}
\centering
\includegraphics[width=0.6\linewidth]{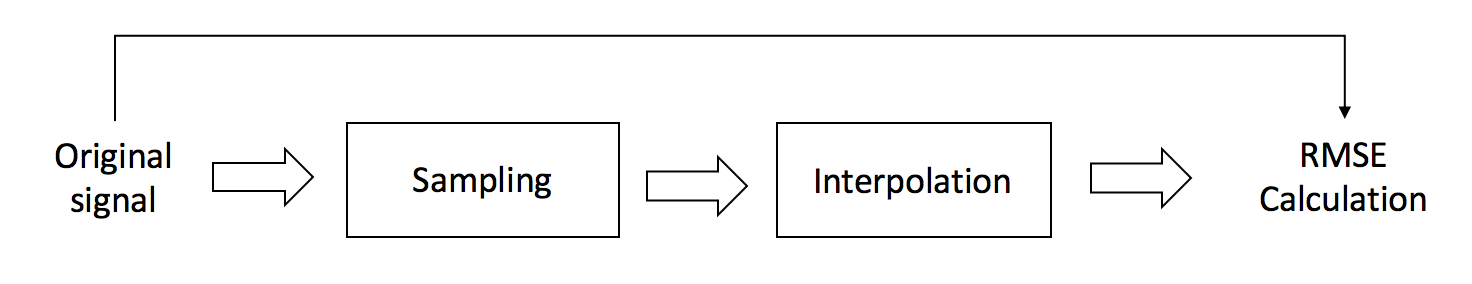}
\caption{The methodology to calculate the best interpolation method is based on the RMSE between the original signal and the reconstructed one.}
\label{fig:Methodology}
\end{figure}

To perform the experiments we followed the methodology explained in Figure \ref{fig:Methodology}. First, we downsampled the original time series (using Lebesgue or Riemann sampling), then we reconstructed the original signals from the downsampled data (using Linear, PCHIP, ZOH... interpolation methods) and finally, we compared the original signal with the reconstructed one (using RMSE) to evaluate the performance of the different interpolation methods.

There are many metrics in the state-of-the-art to calculate the difference between the original signal and the reconstructed signal. In this research, we applied a very popular metric called root-mean-square-error (RMSE). The RMSE has been used in many research works to calculate the efficiency of the interpolation techniques \cite{vzukovivc2008environmental,muhlenstadt2011kernel}. To this end, all the signals of all the datasets have been individually normalised between 0 and 1.

In order to conduct the experiments we applied some of the most popular time series interpolation methods in the state of the art such as ZOH \cite{de1978practical}, Linear interpolation \cite{de1978practical}, PCHIP \cite{kahaner1989numerical}, Shannon \cite{marks2012introduction}, Lasso \cite{tibshirani2011regression}, Natural Neighbour \cite{boissonnat2002smooth}, Cubic \cite{keys1981cubic}, Multiquadric \cite{hon1997multiquadric}, Inverse Multiquadric \cite{buhmann1992multiquadric}, Gaussian \cite{harville1974bayesian}, Quintic \cite{erkorkmaz2005quintic} and Thin-Plate \cite{bookstein1989principal}. This functions have been implemented using the Radial Basis Function (RBF) approximation/interpolation in python based on the books\cite{fasshauer2007meshfree} and \cite{schimek2013smoothing}. 

\bigbreak

We have applied two different approaches strategies to downsample the signals:

\begin{itemize}
\item Lebesgue sampling: Our implementation of Lebesgue sampling is based on the absolute difference between the sampled values. In other words, the current value of the signal is captured when its difference from the last sampled point exceeds a preset limit.
\item Riemann sampling: The Riemann sampling is performed by using the same (or slightly higher) average number of points than in Lebesgue sampling but using a fix time interval over time. Riemann sampling always takes the same or more points than with Lebesgue sampling because we adapted the threshold of Lebesgue sampling for not taking a higher percentage of samples than the established limit, so normally the percentage is slightly lower.
\end{itemize}

To perform the experiments all the signals of all the datasets have been interpolated between 0 and 1. The values of the parameters for the developed methods (ZeLi, ZeLiC, ZeChip, and ZeChipC) were: \textit{tolerance ratio = 1.15, min distance=3, and previous distance=3}. As shown in Table \ref{TableI}, of the appendix depending on the dataset more or fewer samples were selected. 

The objective of the first experiment is to evaluate the performance of our proposed method for interpolating time series from Lebesgue sampling. To this end, we compared our methods with those of the state-of-the-art. In this experiment, we applied Lebesgue sampling based on the difference between the values with a threshold of 0.05\footnote{Note that all the signals had been scaled between 0 and 1. From this perspective 0.05 means a 5\% difference in the maximum change between the possible values.}.

The goal of the second experiment is a bit more ambitious than the one of the first experiment. We want to demonstrate that the best technique to interpolate and reconstruct any signal is to use Lebesgue sampling and our best-proposed method; ZeChipC. To this end, we will conduct a similar experiment to the first one but this time with the same number of samples for both Lebesgue sampling and Riemann sampling. We will select 15\% of the total samples of the signal for both Riemann and Lebesgue sampling. In the case of Lebesgue, we will tune the value of the threshold until selecting the same or slightly less (never more) number of samples than in Riemann sampling.

To carry out the experiments we used Python 3.6.1 with Anaconda custom (x86\_64). We used a MacBook with macOS High Sierra with the following features. 2.3 GHz Intel Core i5, 8GB 2133 MHz, L2 Cache:256 KB, L3: 4MB.

\subsection{Data sets}

As described in Table \ref{TableI} of the appendix, the experiments have been conducted over the 67 databases of a repository called ``The UEA and UCR time series classification repository" \cite{bagnall2018uea}. Some of the datasets have the training set and the testing set to perform classification techniques. Since we are not doing classification, we have simply joined the training set and testing set in a single dataset for each dataset of the repository. 

\subsection{Experiment I}

The results of this experiment for all the 16 methods are found in the appendix section in Table \ref{TableII}. Figure \ref{fig:Exp1_AvgRMSE} shows an illustrative summary of the performance of the best eight combinations of interpolation methods and sampling strategies. In Table \ref{tab:top-results-exp1} it can be seen the position of the ranking and the average RMSE error of the 67 datasets.

\begin{table}[htbp]
\caption{Average RMSE of the top 10 methods in Lebesgue sampling}
\centering { \noindent\adjustbox{max width=0.95\textwidth}{%
\begin{tabular}{|l|c|c|c|c|c|c|c|c|c|c|}
\hline
Method & L ZeChipC & L ZeLiC & L ZeChip & L ZeLi & L Zero & L PCHIP & L Linear & L Nearest & L Shannon & L Thin-Plate \\ \hline
Position & 1 & 2 & 3 & 4 & 5 & 6 & 7 & 8 & 9 & 10 \\ \hline
Avg RMSE & \textbf{0.0029} & 0.0030 & 0.0031 & 0.0034 & \textbf{0.0040} & 0.0049 & 0.0055 & 0.0067 & 0.0406 & 0.0554 \\ \hline
\end{tabular}
\label{tab:top-results-exp1}
}}
\end{table}

\begin{figure}
\centering
\includegraphics[width=0.75\linewidth]{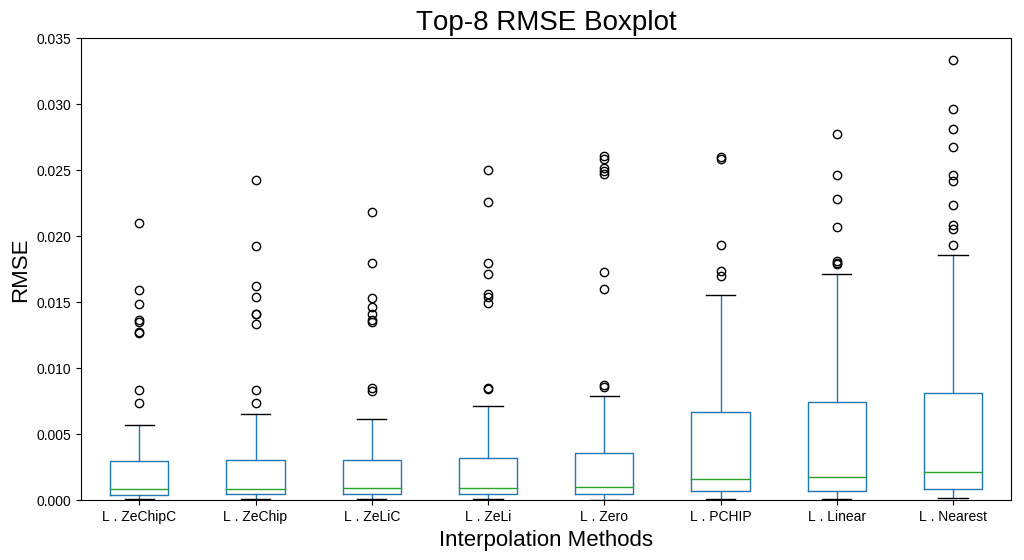}
\caption{Boxplot of the RMSE of the top-8 methods for all the 67 datasets, ordered by the median value.}
\label{fig:Exp1_AvgRMSE}
\end{figure}

\begin{figure}
\centering
\includegraphics[width=0.75\linewidth]{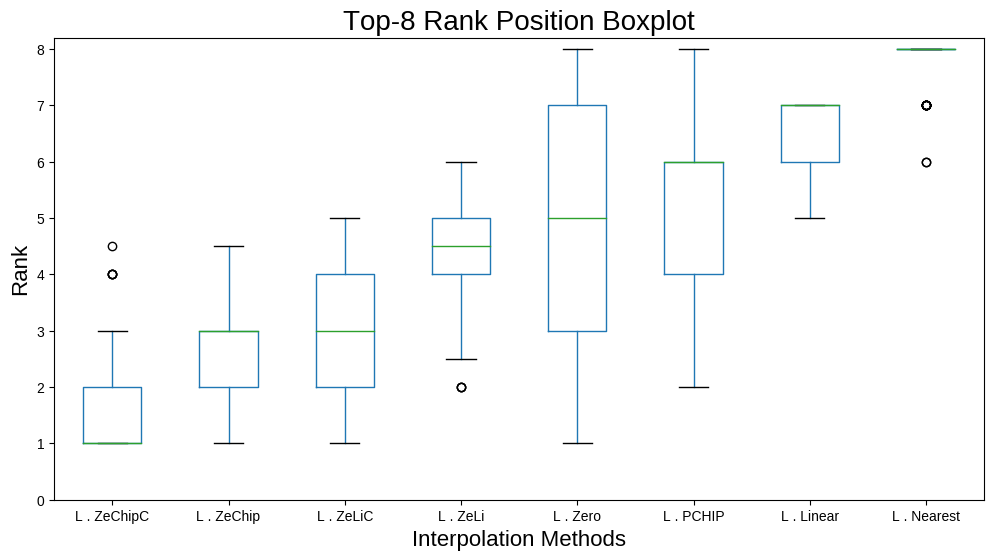}
\caption{Boxplot of the rank position of the top-8 methods for all the 67 datasets.}
\label{fig:RankingPositionExp}
\end{figure}

It could be possible that a method is strongly penalised in some dataset and this will undermine its performance severely. To avoid this, we also calculated the average position as displayed in Figure \ref{fig:RankingPositionExp}. We can see that the position order of the median RMSE is the same as the order in terms of the average position of the Ranking.

Figure \ref{fig:Exp1_AvgRMSE} shows the box-plot of the RMSE score of the top-8 interpolation methods. In this plot, the interpolation methods are sorted by the 50th percentile, and from the figure, it can be seen that ZeChipC performs best for the Lebesgue sampling and it produces least errors while reconstructing the signal. Furthermore, it can be observed that interquartile range is smaller in magnitude compared to the rest of methods and as well whisker is at lower RMSE value, which establishes that ZeChipC performs very well in the overall spread of the reconstruction from the sampled single.

Likewise, Figure \ref{fig:RankingPositionExp} shows a similar conclusion where ZeChipC is the winner in terms of the rank position of the reconstructed signal. The 50th percentile shows that ZeChipC half of the time is a clear winner compared to the rest of the interpolation methods. Similarly,  the interquartile range and the whisker establishes that overall ZeChipC outperforms the rest of the interpolation methods.

\subsection{Experiment II}

\begin{figure}
\centering
\includegraphics[width=0.75\linewidth]{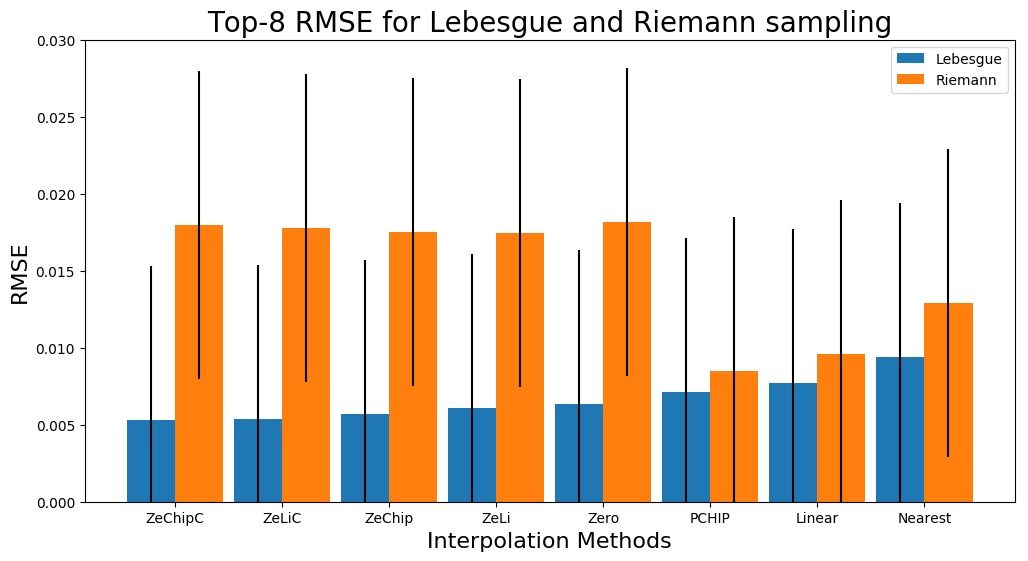}
\caption{RMSE of the top-8 methods for all the 67 datasets.}
\label{fig:Exp2_AvgRMSE}
\end{figure}

In this experiment, results are shown in the same way as in the first one. First, in Figure \ref{fig:Exp2_AvgRMSE} it is shown the performance based on the average RMSE of each combination (sampling and interpolation methods) using 15\% of the samples for each of the 67 datasets. As shown in Table \ref{TableI}, we adapted the threshold in Lebesgue sampling not to collect more than 15\% for each dataset as it can be seen in Table \ref{TableIII} of the appendix.  In Table \ref{tab:top-results-exp2} it can be seen the position of the ranking and the median RMSE error of the 67 datasets.

\begin{table}[htbp]
\caption{Average RMSE of the top 12 methods using Lebesgue and Riemann sampling}
\centering { \noindent\adjustbox{max width=1\textwidth}{%
\begin{tabular}{|c|c|c|c|c|c|c|c|c|c|c|c|c|}
\hline
Method & L ZeChipC & L ZeLiC & L ZeChip & L ZeLi & L Zero & L PCHIP & L Linear & R PCHIP & R Quintic & R Cubic & L Nearest & R Thin-Plate \\ \hline
Position & 1 & 2 & 3 & 4 & 5 & 6 & 7 & 8 & 9 & 10 & 11 & 12 \\ \hline
Avg RMSE & \textbf{0.0053} & 0.0054 & 0.0057 & 0.0061 & \textbf{0.0063} & 0.0071 & 0.0077 & 0.0085 & 0.0087 & 0.0091 & 0.0094 & 0.0095 \\ \hline
\end{tabular}
\label{tab:top-results-exp2}
}}
\end{table}

\begin{figure}
\centering
\includegraphics[width=0.75\linewidth]{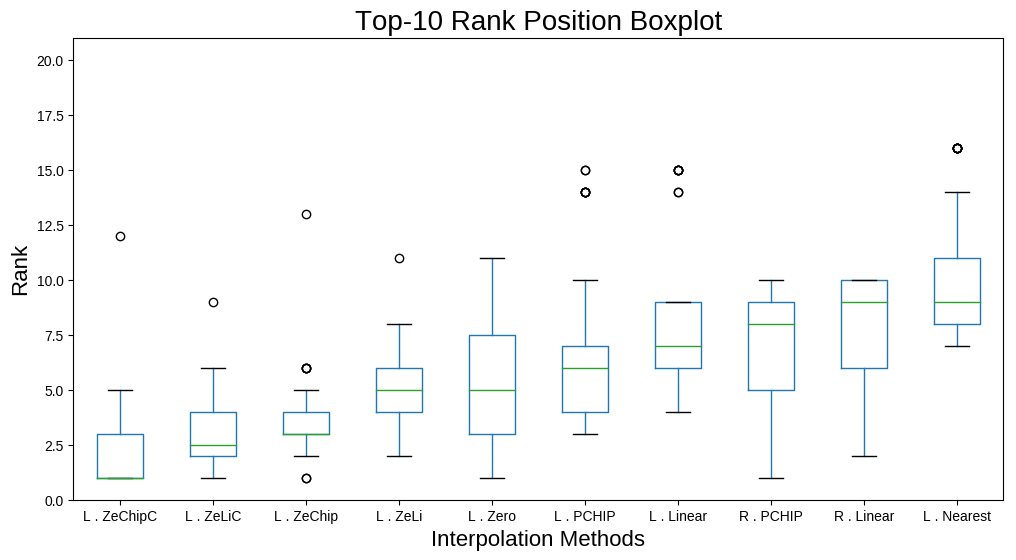}
\caption{Boxplot of the rank position of the top-10 methods for all the 67 datasets.}
\label{fig:RankingPositionExpII}
\end{figure}

As in the first experiment, in Figure \ref{fig:RankingPositionExpII} it is shown the average position of the best 12 methods of the dataset. We can see that the order of the positions of the Average Ranking is similar to that with the median RMSE value. For example, the order for the first eight combinations is the same.

We can also see that ZeChipC with Lebesgue sampling is the winner in terms of the rank position of the reconstructed signal. The 50th percentile shows that ZeChipC half of the time is the clear winner compared to the rest of the interpolation methods. Similarly,  the interquartile range and the whisker establishes that overall ZeChipC outperforms the rest of the interpolation methods.

\subsection{Discussion of the experiments}

The interpolation method that offers the best performance in both experiments is ZeChipC. This method implements three ideas that have been presented throughout the paper. 

\begin{figure}
\centering
\includegraphics[width=1.0\linewidth]{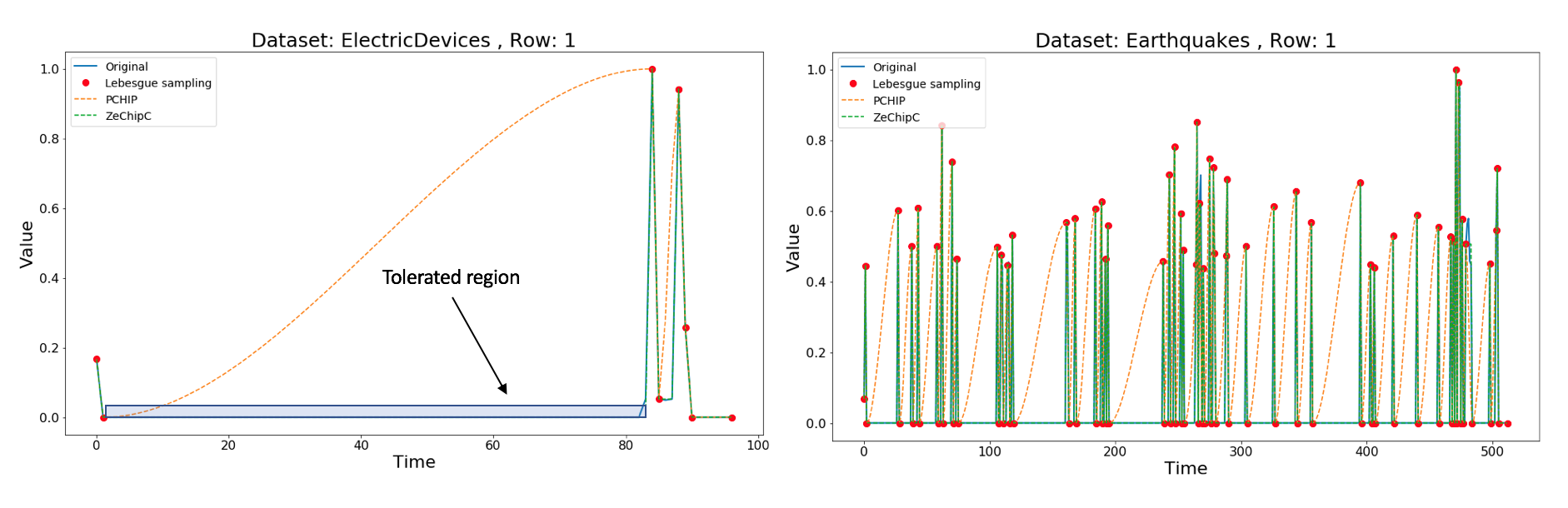}
\caption{(Left) PCHIP is interpolating outside the tolerated region and so, its performance is low. (Right) This can happen several times for the same signal}
\label{fig:PCHIP-Tolerated}
\end{figure}

First, it uses ZOH interpolation which allows ZeChipC to adapt to abrupt changes. This improvement is shared by the other three developed methods (ZeLi, ZeLiC and ZeChip), and it can be clearly appreciated when we compare in experiment I the performance of ZeLi against Linear interpolation or ZeChip against PCHIP interpolation. ZOH is the only interpolation technique that guarantees that all the points are in the tolerated region which allows representing the shape of the signal more accurately. We can see a clear example of this in Figure \ref{fig:PCHIP-Tolerated} (left) where PCHIP interpolation is out of the tolerated region while ZeChip is respecting it.

Second, ZeChipC includes a new functionality to adapt to concave and convex regions (See Figure \ref{fig:Concavity}. It is interesting to see that this improvement means an increment of 8.06\% of ZeChipC with respect ZeChip (which does not implement the convexity/concavity functionality) in the first experiment and of 7.36\% in the second. In the same way, there is an improvement of ZeLiC against ZeLi of 10.88\% and 10.89\% in the first and second experiments respectively.

The third and last idea consists of implementing PCHIP interpolation instead of Linear interpolation. The increase in the performance of this approach can be appreciated when ZeChip is compared against ZeLi, and when ZeChipC is compared against ZeLiC. In the first experiment, ZeChip has an improvement of 8.82\% against ZeLi while in the second experiment it has an improvement of 5.94\%. In the same way, ZeChipC has an improvement against ZeLiC of 5.94\% in the first and of 2.22\% in the second. Something we could ask ourselves is the differences in the performance are statically significant. To this end, we performed the following statistical tests can compare the average performance in both experiments. We compared ZeChip with ZeLi and compared ZeChipC with ZeLiC.

In addition, it is worth stressing that the ZOH interpolation is better than Linear interpolation and even better than PCHIP interpolation. In fact, it is better than any other interpolation technique when Lebesgue sampling is used. PCHIP is the second best and Linear the third best. This is a confirmation that using ZOH for the algorithms ZeChipC and ZeLiC (as well as ZeChip and ZeLi) as a combination of these methods is a good idea. Regarding the rest of the methods, it seems that Lebesgue PCHIP, variable near, variable linear and variable ZOH always remain ahead of the rest of the methods. Our results strengthen the claims regarding that sampling based on Lebesgue sampling is more accurate than Riemann sampling (either in fixed or uniform with the same number of samples).

On the other hand, we could think that ZeChipC is not better because for some data sets simply because it has been in a better average position. To argue for the ZeChipC we have two arguments. In the position ranking, it has won 53 times out of 67 in the first experiment and 37 in the second. And if we see the average position it has been the first method in both experiments, 2.18 in the first experiment and 2.63 in the second.

\begin{figure}
\centering
\includegraphics[width=0.7\linewidth]{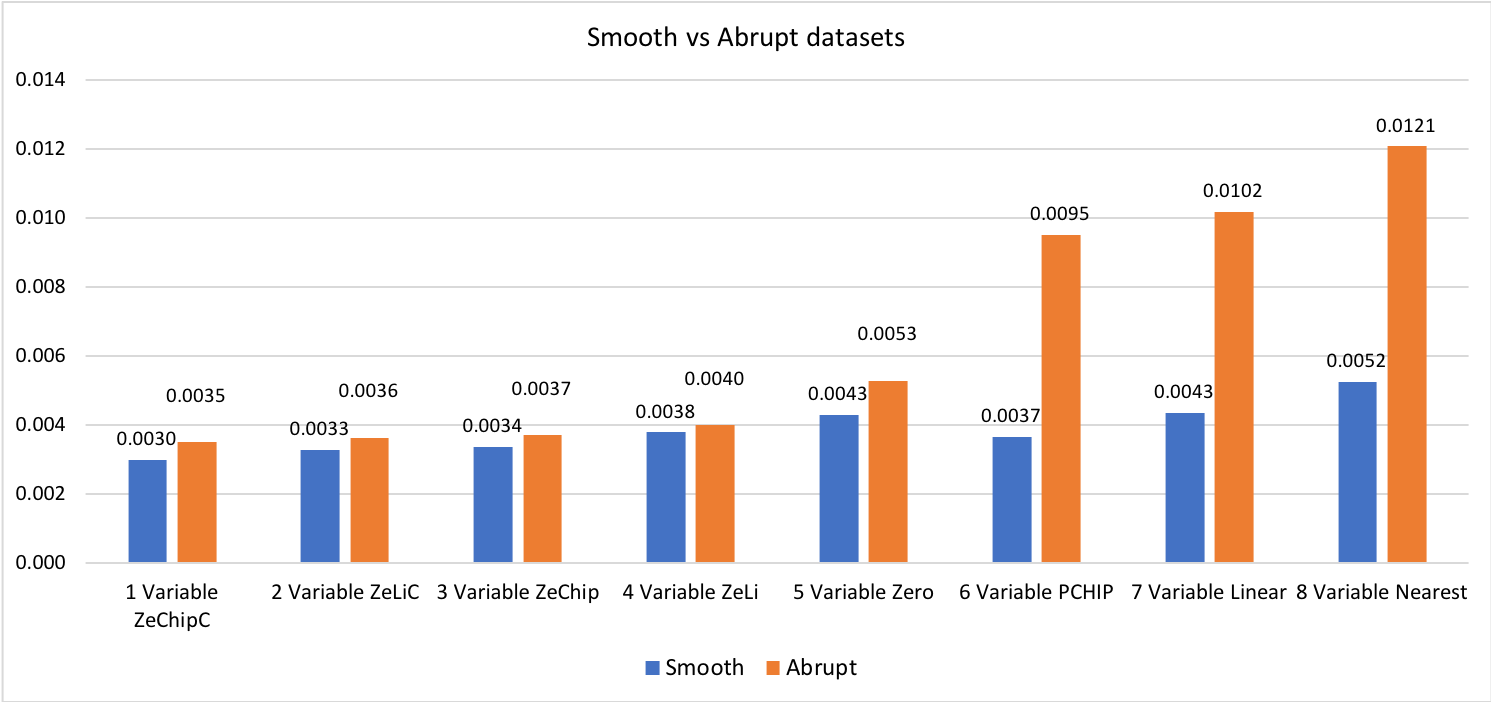}
\caption{Average RMSE of the 15 smoother datasets against the 15 abrupter ones.}
\label{fig:smooth-abrupt}
\end{figure}

Lastly, time series smoothness is a concept that has been studied in detail in several investigations \cite{barnes2003variogram}. One of the most frequent ways of measuring it, and the one applied in our research, is by calculating the standard deviation of the differences between the points (1st derivative). The lower the SD, the smoother the time series is. 

The differences between the proposed methods and the rest of the methods are enlarged when the databases have a large number of changes. As it is shown in Figure \ref{fig:smooth-abrupt}, when the signals of a dataset have abrupt changes, the state-of-the-art methods do not "understand" that the change has occurred between the last tracked point and the previous instant and as a result, the signal is drawn out of the tolerance region as shown in Figure \ref{fig:PCHIP-Tolerated}.

\section{Conclusion}

The main reason why the developed methods (ZeLi, ZeLiC, ZeChip and ZeChipC) have better results than the other interpolation methods is that the interpolation is performed taking into account the Lebesgue sampling characteristics. That is to say, when there is an abrupt change ZOH interpolation is applied, otherwise (when there is a smooth change) Linear and PCHIP interpolation are applied. The proposed methods detect that there has been an abrupt change because the newly captured sample is far away from the tolerated region. Additionally, this decision can be optimised depending on the dataset using the tolerance ratio parameter.

On the other hand, the convexity/concavity functionality has performed very well. We can guess that when there is a change of sign in the slope of a signal, a concavity/convexity region has happened. Optimising the three parameters (previous, minimum forward, and maximum forward) to decide whether or not there is a convex/concave region for each data set as well as establishing a methodology to this end could be a very interesting research path to follow. Additionally, accurately calculating the exact point where the signal changes the slope and approximating its shape is a very complex and wide issue, although the applied implementation performed very well and boosted the performance of both methods: ZeLiC and ZeChipC.

The developed methods have been implemented based on the absolute difference with respect to the last sampled point. However, it is easily adaptable to another kind of events that trigger sensors. For example, the sensor could be triggered when the output signal crosses a certain limit or when the percentage variation is higher than a preset limit. Using the same approach; Linear or PCHIP interpolation for smooth transitions and ZOH for abrupt changes will still be effective.

From the assumptions and contributions of this research, new and more effective interpolation methods could be designed. Lebesgue sampling is known in academia, but it is important to develop reliable and adapted tools to encourage the industry to make a transition to Lebesgue sampling.

\textbf{Acknowledgements.} This publication has emanated from research conducted with the support of Enterprise Ireland (EI), under Grant Number IP20160496 and TC20130013.

\bibliographystyle{apalike} 
\bibliography{references}

\section*{Appendix}

In the appendix section, it is shown the performance of each method for the whole dataset. In the Table \ref{tab:information-dataset} , we show the following fields. Number of columns and number of rows, metric to measure how abrupt the dataset is, Value of the threshold, Average number of samples for the first experiment, the value of the threshold to capture 15\% of the samples.

\begin{table}[htbp] 
\caption{Information about the datasets and the experiments.}\label{tab:information-dataset}
\centering { \noindent\adjustbox{max width=0.8\textwidth}{%
\begin{tabular}{|c|l|c|c|c|c|c|c|c|}
\hline
N & Data set & Rows & Columns & Abrupt & Exp I: Thres & Exp I: Perc & Exp II: Thres & Exp II: Perc \\ \hline
1 & Adiac & 779 & 177 & 11.12 & 0.05 & 9.81 & 0.0305 & 14.93 \\ \hline
2 & ArrowHead & 209 & 252 & 12.71 & 0.05 & 22.53 & 0.0794 & 14.97 \\ \hline
3 & Beef & 58 & 471 & 14.46 & 0.05 & 8.8 & 0.0287 & 14.99 \\ \hline
4 & BeetleFly & 38 & 513 & 149.5 & 0.05 & 31.21 & 0.1041 & 15 \\ \hline
5 & BirdChicken & 38 & 513 & 135.95 & 0.05 & 14.52 & 0.0483 & 14.94 \\ \hline
6 & CBF & 928 & 129 & 14.31 & 0.05 & 72.79 & 0.2205 & 14.77 \\ \hline
7 & Car & 118 & 578 & 24.06 & 0.05 & 9.57 & 0.0312 & 14.95 \\ \hline
8 & ChlorineConcentration & 4305 & 167 & 102.34 & 0.05 & 51.58 & 0.2256 & 14.94 \\ \hline
9 & Coffee & 54 & 287 & 11.04 & 0.05 & 19.9 & 0.0664 & 14.98 \\ \hline
10 & Computers & 498 & 721 & 303.2 & 0.05 & 13.77 & 0.0498 & 14.45 \\ \hline
11 & DiatomSizeReduction & 320 & 346 & 27.74 & 0.05 & 15.38 & 0.0511 & 15 \\ \hline
12 & DistalPhalanxOutlineAgeGroup & 537 & 81 & 7.99 & 0.05 & 63.06 & 0.2427 & 14.98 \\ \hline
13 & DistalPhalanxOutlineCorrect & 874 & 81 & 15.28 & 0.05 & 67.61 & 0.2559 & 14.85 \\ \hline
14 & DistalPhalanxTW & 537 & 81 & 7.08 & 0.05 & 19.01 & 0.0631 & 14.98 \\ \hline
15 & ECG200 & 198 & 97 & 50.06 & 0.05 & 36.25 & 0.1384 & 14.95 \\ \hline
16 & ECG5000 & 4998 & 141 & 42.17 & 0.05 & 20.23 & 0.0744 & 14.97 \\ \hline
17 & ECGFiveDays & 882 & 137 & 30.57 & 0.05 & 15.16 & 0.0513 & 14.98 \\ \hline
18 & Earthquakes & 459 & 513 & 656.47 & 0.05 & 28.9 & 0.5675 & 15 \\ \hline
19 & ElectricDevices & 16635 & 97 & 37.83 & 0.05 & 9.2 & 0 & 11.67 \\ \hline
20 & FaceAll & 2248 & 132 & 12.39 & 0.05 & 61.3 & 0.238 & 14.99 \\ \hline
21 & FaceFour & 110 & 351 & 178.73 & 0.05 & 24.61 & 0.1003 & 14.98 \\ \hline
22 & FacesUCR & 2248 & 132 & 13.84 & 0.05 & 43.34 & 0.1597 & 14.99 \\ \hline
23 & Fish & 348 & 464 & 33.92 & 0.05 & 9.7 & 0.0323 & 14.96 \\ \hline
24 & FordA & 4919 & 501 & 352.36 & 0.05 & 49.95 & 0.172 & 14.98 \\ \hline
25 & FordB & 4444 & 501 & 449.73 & 0.05 & 49.24 & 0.1644 & 14.97 \\ \hline
26 & Ham & 212 & 432 & 34.3 & 0.05 & 24.25 & 0.0852 & 14.99 \\ \hline
27 & HandOutlines & 1368 & 2710 & 39.08 & 0.05 & 3.1 & 0.0102 & 14.85 \\ \hline
28 & Haptics & 461 & 1093 & 20.63 & 0.05 & 3.4 & 0.0089 & 14.92 \\ \hline
29 & Herring & 126 & 513 & 1688.73 & 0.05 & 14.17 & 0.0472 & 14.94 \\ \hline
30 & InlineSkate & 648 & 1883 & 44.37 & 0.05 & 1.85 & 0.0067 & 14.7 \\ \hline
31 & InsectWingbeatSound & 2198 & 257 & 15.12 & 0.05 & 13.81 & 0.0448 & 14.98 \\ \hline
32 & ItalyPowerDemand & 1094 & 25 & 7.03 & 0.05 & 67.16 & 0.6175 & 14.95 \\ \hline
33 & LargeKitchenAppliances & 748 & 721 & 107.06 & 0.05 & 3.98 & 0.0011 & 13.78 \\ \hline
34 & Mallat & 2398 & 1025 & 25.6 & 0.05 & 7.65 & 0.0255 & 14.93 \\ \hline
35 & Meat & 118 & 449 & 12.13 & 0.05 & 7.13 & 0.0225 & 14.97 \\ \hline
36 & MedicalImages & 1139 & 100 & 8.26 & 0.05 & 16.37 & 0.0564 & 14.95 \\ \hline
37 & MiddlePhalanxOutlineAgeGroup & 552 & 81 & 7.5 & 0.05 & 56.63 & 0.207 & 14.98 \\ \hline
38 & MiddlePhalanxOutlineCorrect & 889 & 81 & 10.75 & 0.05 & 61.37 & 0.2272 & 14.98 \\ \hline
39 & MiddlePhalanxTW & 551 & 81 & 6.86 & 0.05 & 20.57 & 0.0705 & 14.98 \\ \hline
40 & MoteStrain & 1270 & 85 & 43.66 & 0.05 & 31.24 & 0.1458 & 14.98 \\ \hline
41 & OSULeaf & 440 & 428 & 56.13 & 0.05 & 19.87 & 0.0667 & 14.92 \\ \hline
42 & OliveOil & 58 & 571 & 19.73 & 0.05 & 10.76 & 0.0334 & 14.98 \\ \hline
43 & PhalangesOutlinesCorrect & 2656 & 81 & 11.86 & 0.05 & 67.61 & 0.2559 & 14.85 \\ \hline
44 & Phoneme & 2108 & 1025 & 50.35 & 0.05 & 30.43 & 0.1072 & 15 \\ \hline
45 & Plane & 208 & 145 & 20.24 & 0.05 & 35.43 & 0.1227 & 14.99 \\ \hline
46 & ProximalPhalanxOutlineAgeGroup & 603 & 81 & 7.61 & 0.05 & 58.35 & 0.2245 & 14.98 \\ \hline
47 & ProximalPhalanxOutlineCorrect & 889 & 81 & 9.62 & 0.05 & 61.92 & 0.2511 & 14.94 \\ \hline
48 & ProximalPhalanxTW & 603 & 81 & 6.9 & 0.05 & 27.81 & 0.1045 & 14.94 \\ \hline
49 & RefrigerationDevices & 748 & 721 & 342.14 & 0.05 & 24.9 & 0.1111 & 11.2 \\ \hline
50 & ScreenType & 748 & 721 & 107.37 & 0.05 & 18.69 & 0.0555 & 13.95 \\ \hline
51 & ShapeletSim & 198 & 501 & 788.94 & 0.05 & 89.69 & 0.5078 & 14.67 \\ \hline
52 & ShapesAll & 1198 & 513 & 22.77 & 0.05 & 11.24 & 0.0367 & 14.98 \\ \hline
53 & SmallKitchenAppliances & 748 & 721 & 181.97 & 0.05 & 2.77 & 0 & 3.61 \\ \hline
54 & StarlightCurves & 9234 & 1025 & 26.36 & 0.05 & 3.7 & 0.0116 & 14.96 \\ \hline
55 & Strawberry & 981 & 236 & 10.78 & 0.05 & 18.62 & 0.0636 & 14.96 \\ \hline
56 & SwedishLeaf & 1123 & 129 & 19.7 & 0.05 & 22.68 & 0.0769 & 14.97 \\ \hline
57 & Symbols & 1018 & 399 & 50.19 & 0.05 & 7.75 & 0.0241 & 14.97 \\ \hline
58 & ToeSegmentation1 & 266 & 278 & 164.52 & 0.05 & 26.67 & 0.0945 & 14.98 \\ \hline
59 & ToeSegmentation2 & 164 & 344 & 164.01 & 0.05 & 20.95 & 0.07 & 14.95 \\ \hline
60 & Trace & 198 & 276 & 37.55 & 0.05 & 5.12 & 0.0161 & 14.88 \\ \hline
61 & TwoLeadECG & 1160 & 83 & 11.83 & 0.05 & 30.79 & 0.1219 & 14.97 \\ \hline
62 & UWaveGestureLibraryAll & 4476 & 946 & 53.56 & 0.05 & 8.27 & 0.0275 & 14.97 \\ \hline
63 & Wafer & 7162 & 153 & 297.74 & 0.05 & 8.46 & 0.0116 & 14.6 \\ \hline
64 & Wine & 109 & 235 & 9.52 & 0.05 & 19.29 & 0.0628 & 14.96 \\ \hline
65 & Worms & 256 & 901 & 72.27 & 0.05 & 13.17 & 0.0436 & 14.95 \\ \hline
66 & WormsTwoClass & 256 & 901 & 79.41 & 0.05 & 13.17 & 0.0436 & 14.95 \\ \hline
67 & Yoga & 3298 & 427 & 1217.33 & 0.05 & 15.55 & 0.0519 & 14.95 \\ \hline
\end{tabular}

}}
\label{TableI}
\end{table}

\begin{table}[htbp]
\caption{Experiment I: RMSE per data set of the top 12 methods using Lebesgue sampling with 0.05 Threshold.}
\centering { \noindent\adjustbox{max width=\textwidth}{%
\begin{tabular}{|c|c|c|c|c|c|c|c|c|c|c|c|c|c|c|c|c|}
\hline
\textbf{N} & \textbf{ZeChipC} & \textbf{ZeLiC} & \textbf{ZeChip} & \textbf{ZeLi} & \textbf{Zero} & \textbf{PCHIP} & \textbf{Linear} & \textbf{Nrst} & \textbf{Shannon} & \textbf{T-P} & \textbf{Lasso} & \textbf{Cubic} & \textbf{Quintic} & \textbf{Inv-mlt} & \textbf{Mltqdc} & \textbf{Gaussian} \\ \hline
1 & \textbf{0.0159} & 0.018 & 0.0192 & 0.0226 & 0.0261 & 0.0193 & 0.0228 & 0.0268 & 0.2397 & 0.2737 & 0.1613 & 0.7989 & 8.433 & 0.7879 & 1.078 & 1.165 \\ \hline
2 & \textbf{0.0148} & 0.0153 & 0.0162 & 0.018 & 0.0249 & 0.0169 & 0.0206 & 0.0281 & 0.21 & 0.0629 & 0.6052 & 0.0963 & 0.0957 & 0.1292 & 0.1119 & 0.4174 \\ \hline
3 & \textbf{0.0209} & 0.0218 & 0.0242 & 0.025 & 0.0247 & 0.0258 & 0.0277 & 0.0333 & 0.2555 & 0.3284 & 0.4262 & 0.9974 & 2.568 & 0.5583 & 2.052 & 74442.7 \\ \hline
4 & \textbf{0.0135} & 0.0146 & 0.0141 & 0.0156 & 0.0251 & 0.0139 & 0.0171 & 0.0241 & 0.1583 & 0.0196 & 0.5401 & 0.0212 & 0.0244 & 0.0328 & 0.0208 & 0.1151 \\ \hline
5 & 0.0137 & 0.0141 & 0.0154 & 0.0171 & \textbf{0.0258} & 0.0152 & 0.0179 & 0.0246 & 0.1714 & 0.0243 & 0.5603 & 0.0384 & 0.0486 & 0.0571 & 0.0385 & 0.189 \\ \hline
6 & \textbf{0.0008} & 0.0008 & 0.0008 & 0.0008 & 0.0006 & 0.0018 & 0.0017 & 0.0016 & 0.0066 & 0.0044 & 0.0209 & 0.0049 & 0.0064 & 0.0045 & 0.005 & 0.006 \\ \hline
7 & \textbf{0.0041} & \textbf{0.0044} & \textbf{0.0047} & \textbf{0.0055} & \textbf{0.0085} & 0.0047 & 0.0055 & 0.0077 & 0.0405 & 0.0096 & 0.1527 & 0.0171 & 0.0223 & 0.0277 & 0.0243 & 0.0593 \\ \hline
8 & \textbf{0.0002} & 0.0002 & 0.0002 & 0.0002 & 0.0002 & 0.0007 & 0.0007 & 0.0008 & 0.0018 & 0.0012 & 0.0025 & 0.0016 & 0.0039 & 0.0011 & 0.0015 & 0.0015 \\ \hline
9 & 0.0127 & 0.0136 & 0.0141 & 0.0154 & \textbf{0.0173} & 0.0155 & 0.018 & 0.0223 & 0.143 & 0.0215 & 0.3963 & 0.0382 & 0.1034 & 0.0372 & 0.0392 & 2.14 \\ \hline
10 & \textbf{0.0012} & 0.0012 & 0.0012 & 0.0012 & 0.001 & 0.0147 & 0.0171 & 0.0208 & 0.0456 & 0.3657 & 0.0222 & 1.906 & 683.2 & 3125.2 & 13844.8 & 1763246.9 \\ \hline
11 & \textbf{0.0013} & 0.0016 & \textbf{0.0016} & 0.002 & 0.0032 & 0.0016 & 0.0021 & 0.0028 & 0.0138 & 0.0029 & 0.0495 & 0.0043 & 0.0041 & 0.0072 & 0.0053 & 0.0102 \\ \hline
12 & \textbf{0.0012} & 0.0013 & \textbf{0.0012} & 0.0013 & \textbf{0.0013} & 0.0014 & 0.0018 & 0.002 & 0.0098 & 0.0032 & 0.035 & 0.0035 & 0.0036 & 0.0056 & 0.0032 & 0.0148 \\ \hline
13 & \textbf{0.0007} & 0.0008 & \textbf{0.0007} & 0.0008 & 0.0007 & 0.0008 & 0.001 & 0.0012 & 0.0064 & 0.002 & 0.0248 & 0.0021 & 0.0022 & 0.0033 & 0.002 & 0.0092 \\ \hline
14 & \textbf{0.0011} & 0.0012 & 0.0011 & 0.0012 & \textbf{0.0013} & 0.0012 & 0.0016 & 0.0019 & 0.0179 & 0.0133 & 0.0139 & 0.0477 & 0.4118 & 0.0064 & 0.0104 & 0.0137 \\ \hline
15 & \textbf{0.0041} & \textbf{0.0042} & \textbf{0.0042} & \textbf{0.0042} & \textbf{0.0041} & \textbf{0.0058} & 0.0058 & 0.0068 & 0.0496 & 0.0096 & 0.091 & 0.0126 & 0.0297 & 0.0142 & 0.0136 & 0.0736 \\ \hline
16 & \textbf{0.0002} & \textbf{0.0002} & \textbf{0.0002} & \textbf{0.0002} & 0.0002 & 0.0002 & 0.0003 & 0.0003 & 0.002 & 0.0004 & 0.0052 & 0.0009 & 0.0033 & 0.0011 & 0.0031 & 0.1478 \\ \hline
17 & \textbf{0.0007} & \textbf{0.0007} & \textbf{0.0007} & \textbf{0.0007} & \textbf{0.001} & 0.0019 & 0.0021 & 0.0025 & 0.0112 & 0.0097 & 0.0297 & 0.0273 & 0.3053 & 0.0875 & 0.4517 & 38.8 \\ \hline
18 & \textbf{0.0003} & \textbf{0.0003} & \textbf{0.0003} & \textbf{0.0003} & \textbf{0.0003} & 0.026 & 0.0246 & 0.0296 & 0.0377 & 0.0488 & 0.0208 & 0.1096 & 1.14 & 0.6394 & 1.821 & 47.9 \\ \hline
19 & \textbf{0} & 0 & \textbf{0} & 0 & \textbf{0} & 0.0006 & 0.0006 & 0.0007 & 0.0011 & 0.0045 & 0.0003 & 0.0166 & 1.112 & 0.0778 & 0.3465 & 1.149 \\ \hline
20 & 0.0003 & 0.0004 & 0.0003 & 0.0004 & \textbf{0.0003} & 0.0004 & 0.0005 & 0.0005 & 0.0043 & 0.0008 & 0.0089 & 0.0008 & 0.001 & 0.0009 & 0.0008 & 0.0023 \\ \hline
21 & 0.0083 & 0.0083 & 0.0083 & 0.0084 & \textbf{0.0068} & 0.0173 & 0.0181 & 0.0193 & 0.0891 & 0.0272 & 0.1725 & 0.0422 & 0.1634 & 0.0349 & 0.0506 & 0.2587 \\ \hline
22 & \textbf{0.0004} & 0.0004 & 0.0004 & 0.0004 & 0.0003 & 0.0005 & 0.0006 & 0.0006 & 0.0033 & 0.0013 & 0.0059 & 0.0018 & 0.0041 & 0.0015 & 0.0018 & 0.003 \\ \hline
23 & \textbf{0.0016} & 0.0017 & \textbf{0.0018} & 0.002 & 0.0029 & 0.0018 & 0.002 & 0.0027 & 0.0127 & 0.0049 & 0.0393 & 0.0081 & 0.0084 & 0.0119 & 0.0102 & 0.0187 \\ \hline
24 & \textbf{0.0001} & \textbf{0.0002} & \textbf{0.0001} & \textbf{0.0002} & \textbf{0.0002} & \textbf{0.0002} & \textbf{0.0002} & 0.0003 & 0.002 & 0.0003 & 0.0041 & 0.0003 & 0.0003 & 0.0004 & 0.0003 & 0.0014 \\ \hline
25 & \textbf{0.0002} & 0.0002 & 0.0002 & 0.0002 & 0.0002 & 0.0002 & 0.0002 & 0.0003 & 0.0021 & 0.0003 & 0.0046 & 0.0003 & 0.0003 & 0.0005 & 0.0003 & 0.0015 \\ \hline
26 & \textbf{0.0038} & \textbf{0.0042} & 0.0039 & 0.0044 & 0.0041 & 0.0049 & 0.006 & 0.007 & 0.0549 & 0.0326 & 0.0509 & 0.065 & 0.0666 & 0.0743 & 0.4342 & 618.3 \\ \hline
27 & \textbf{0.0002} & 0.0002 & 0.0003 & 0.0004 & 0.0008 & 0.0003 & 0.0004 & 0.0006 & 0.0027 & 0.0133 & 0.0178 & 0.0305 & 0.0288 & 0.0242 & 0.033 & 0.0273 \\ \hline
28 & \textbf{0.0015} & 0.0016 & 0.0019 & 0.0021 & 0.0021 & 0.002 & 0.0022 & 0.0025 & 0.0159 & 0.0211 & 0.0575 & 0.0406 & 1.867 & 531.1 & 5193.4 & 19604 \\ \hline
29 & \textbf{0.0039} & 0.0041 & 0.0045 & 0.0052 & 0.0079 & 0.0046 & 0.0052 & 0.0072 & 0.0467 & 0.0053 & 0.1642 & 0.0063 & 0.0071 & 0.0112 & 0.0068 & 0.0406 \\ \hline
30 & \textbf{0.0011} & \textbf{0.0012} & \textbf{0.0012} & 0.0014 & \textbf{0.0015} & 0.0012 & 0.0014 & 0.0017 & 0.0152 & 0.0563 & 0.0196 & 0.2342 & 4.325 & 0.5209 & 1.464 & 1745.2 \\ \hline
31 & 0.0004 & 0.0004 & 0.0004 & 0.0005 & \textbf{0.0004} & 0.0005 & 0.0006 & 0.0007 & 0.0056 & 0.008 & 0.0031 & 0.0216 & 0.2183 & 0.0103 & 0.0423 & 418 \\ \hline
32 & \textbf{0.0006} & 0.0006 & \textbf{0.0006} & 0.0006 & 0.0005 & 0.0015 & 0.0016 & 0.0017 & 0.0054 & 0.0033 & 0.0208 & 0.0036 & 0.005 & 0.0034 & 0.0035 & 0.0057 \\ \hline
33 & \textbf{0.0003} & \textbf{0.0004} & \textbf{0.0003} & 0.0004 & 0.0005 & 0.0072 & 0.0093 & 0.0113 & 0.0222 & 0.1143 & 0.0058 & 0.7677 & 155.5 & 9904 & 25632 & 2343150.8 \\ \hline
34 & 0.0003 & 0.0003 & 0.0003 & 0.0004 & \textbf{0.0004} & 0.0004 & 0.0004 & 0.0005 & 0.0033 & 0.0061 & 0.005 & 0.0123 & 0.0168 & 0.0723 & 0.2349 & 139.5 \\ \hline
35 & \textbf{0.0073} & \textbf{0.0085} & \textbf{0.0073} & \textbf{0.0085} & 0.0069 & 0.0075 & 0.0089 & 0.0109 & 0.1179 & 0.9908 & 0.1191 & 1.964 & 1.358 & 0.1303 & 0.5857 & 4794.5 \\ \hline
36 & \textbf{0.0006} & \textbf{0.0006} & \textbf{0.0006} & 0.0006 & 0.0007 & 0.0007 & 0.0007 & 0.0008 & 0.011 & 0.024 & 0.0066 & 0.1826 & 30.6 & 0.023 & 0.1037 & 5.957 \\ \hline
37 & \textbf{0.0012} & 0.0012 & \textbf{0.0012} & 0.0013 & 0.0014 & 0.0014 & 0.0017 & 0.0021 & 0.012 & 0.0027 & 0.0346 & 0.0031 & 0.0032 & 0.0051 & 0.0029 & 0.0142 \\ \hline
38 & \textbf{0.0007} & 0.0008 & \textbf{0.0007} & 0.0008 & 0.0008 & 0.0009 & 0.0011 & 0.0013 & 0.008 & 0.0017 & 0.0246 & 0.0019 & 0.002 & 0.0035 & 0.0017 & 0.0103 \\ \hline
39 & \textbf{0.0011} & \textbf{0.0013} & 0.0011 & 0.0013 & 0.0013 & 0.0012 & 0.0015 & 0.0019 & 0.0167 & 0.0121 & 0.0142 & 0.0422 & 0.3532 & 0.0062 & 0.0097 & 0.0132 \\ \hline
40 & \textbf{0.0005} & \textbf{0.0005} & 0.0006 & 0.0006 & 0.0007 & 0.0025 & 0.0025 & 0.0029 & 0.0076 & 0.0056 & 0.0181 & 0.0107 & 0.039 & 0.0092 & 0.0162 & 0.0226 \\ \hline
41 & \textbf{0.0014} & 0.0014 & 0.0015 & 0.0016 & 0.0022 & 0.0015 & 0.0017 & 0.0022 & 0.0133 & 0.0032 & 0.0378 & 0.0048 & 0.0053 & 0.0047 & 0.0052 & 0.0115 \\ \hline
42 & \textbf{0.0127} & 0.0135 & \textbf{0.0133} & 0.0149 & \textbf{0.016} & 0.0144 & 0.0169 & 0.0205 & 0.2363 & 0.1952 & 0.1931 & 0.6276 & 1.673 & 2.832 & 39.4 & 1286461.6 \\ \hline
43 & \textbf{0.0002} & \textbf{0.0003} & \textbf{0.0002} & \textbf{0.0003} & \textbf{0.0002} & 0.0003 & 0.0003 & 0.0004 & 0.0021 & 0.0007 & 0.0082 & 0.0007 & 0.0007 & 0.0011 & 0.0007 & 0.003 \\ \hline
44 & \textbf{0.0004} & 0.0004 & 0.0004 & 0.0004 & 0.0004 & 0.0005 & 0.0006 & 0.0007 & 0.004 & 0.001 & 0.0097 & 0.0032 & 0.2783 & 0.0017 & 0.0012 & 812.3 \\ \hline
45 & \textbf{0.0031} & 0.0034 & \textbf{0.0032} & 0.0034 & 0.0045 & 0.0035 & 0.0041 & 0.0055 & 0.0256 & 0.0044 & 0.0604 & 0.0044 & 0.0047 & 0.0057 & 0.0045 & 0.0112 \\ \hline
46 & \textbf{0.0011} & \textbf{0.0012} & \textbf{0.0011} & \textbf{0.0012} & 0.0012 & 0.0013 & 0.0017 & 0.002 & 0.0155 & 0.0023 & 0.0308 & 0.0025 & 0.0026 & 0.0048 & 0.0023 & 0.0135 \\ \hline
47 & \textbf{0.0007} & 0.0007 & \textbf{0.0007} & 0.0007 & 0.0008 & 0.0008 & 0.0011 & 0.0013 & 0.0108 & 0.0016 & 0.0235 & 0.0018 & 0.002 & 0.0033 & 0.0016 & 0.0098 \\ \hline
48 & 0.001 & 0.0011 & 0.001 & 0.0012 & \textbf{0.0012} & 0.0011 & 0.0014 & 0.0017 & 0.0149 & 0.008 & 0.0151 & 0.0271 & 0.2449 & 0.0043 & 0.0051 & 0.0084 \\ \hline
49 & 0.0008 & 0.0008 & 0.0008 & 0.0008 & \textbf{0.0007} & 0.0104 & 0.0115 & 0.0137 & 0.0255 & 0.0261 & 0.0239 & 0.0754 & 0.5158 & 0.0354 & 0.1558 & 75.8 \\ \hline
50 & \textbf{0.0005} & \textbf{0.0005} & \textbf{0.0005} & \textbf{0.0005} & \textbf{0.0003} & 0.0069 & 0.0094 & 0.0114 & 0.0293 & 0.3088 & 0.0201 & 1.03 & 87.1 & 105.7 & 4119.2 & 6154367.6 \\ \hline
51 & \textbf{0.0019} & \textbf{0.0019} & 0.0019 & 0.0019 & 0.0019 & 0.0131 & 0.0128 & 0.007 & 0.0721 & 0.049 & 0.1102 & 0.0508 & 0.0552 & 0.0497 & 0.0512 & 0.0516 \\ \hline
52 & \textbf{0.0004} & \textbf{0.0004} & \textbf{0.0005} & \textbf{0.0006} & \textbf{0.0008} & 0.0005 & 0.0006 & 0.0008 & 0.0067 & 0.0011 & 0.0167 & 0.0026 & 0.0075 & 0.0027 & 0.0023 & 0.0072 \\ \hline
53 & \textbf{0.0005} & \textbf{0.0005} & \textbf{0.0005} & \textbf{0.0005} & \textbf{0.0005} & \textbf{0.0149} & \textbf{0.0152} & \textbf{0.0185} & 0.0233 & 0.1386 & 0.0039 & 1.076 & 723.2 & 8955.2 & 10166.5 & 661524.3 \\ \hline
54 & \textbf{0.0001} & 0.0001 & \textbf{0.0001} & 0.0001 & 0.0001 & 0.0001 & 0.0001 & 0.0001 & 0.0008 & 0.0023 & 0.0018 & 0.0057 & 0.0081 & 0.003 & 0.0051 & 0.0215 \\ \hline
55 & \textbf{0.0008} & \textbf{0.0009} & \textbf{0.0008} & 0.0009 & 0.001 & 0.0009 & 0.001 & 0.0012 & 0.0096 & 0.0055 & 0.016 & 0.0092 & 0.0085 & 0.0062 & 0.009 & 0.0333 \\ \hline
56 & \textbf{0.0005} & 0.0005 & 0.0005 & 0.0006 & 0.0008 & 0.0006 & 0.0007 & 0.0009 & 0.0051 & 0.0015 & 0.0082 & 0.0024 & 0.0052 & 0.003 & 0.0038 & 0.0058 \\ \hline
57 & \textbf{0.0004} & \textbf{0.0005} & \textbf{0.0005} & \textbf{0.0006} & 0.001 & 0.0005 & 0.0006 & 0.0009 & 0.0057 & 0.0087 & 0.0162 & 0.0177 & 0.0365 & 0.037 & 0.0637 & 0.2087 \\ \hline
58 & \textbf{0.0031} & 0.0031 & 0.0031 & 0.0031 & 0.0032 & 0.0039 & 0.0042 & 0.0052 & 0.0352 & 0.0076 & 0.0632 & 0.0112 & 0.0412 & 0.0086 & 0.0162 & 0.2704 \\ \hline
59 & \textbf{0.0048} & 0.0049 & 0.0049 & 0.0051 & 0.0053 & 0.0065 & 0.0068 & 0.0084 & 0.0604 & 0.0353 & 0.0978 & 0.0668 & 0.1487 & 0.0207 & 0.0476 & 3.394 \\ \hline
60 & \textbf{0.0024} & \textbf{0.0028} & 0.0025 & 0.0028 & 0.0036 & 0.0036 & 0.0046 & 0.0058 & 0.0524 & 0.3275 & 0.078 & 1.29 & 10.1 & 6.28 & 22.8 & 1832.4 \\ \hline
61 & \textbf{0.0005} & \textbf{0.0005} & \textbf{0.0006} & \textbf{0.0006} & \textbf{0.0007} & 0.0008 & 0.001 & 0.0012 & 0.0075 & 0.0032 & 0.0221 & 0.0064 & 0.0114 & 0.0031 & 0.0044 & 0.0101 \\ \hline
62 & \textbf{0.0002} & \textbf{0.0002} & \textbf{0.0002} & \textbf{0.0002} & \textbf{0.0002} & 0.0003 & 0.0003 & 0.0004 & 0.0017 & 0.0046 & 0.003 & 0.0101 & 0.0132 & 0.0053 & 0.0101 & 305.8 \\ \hline
63 & \textbf{0.0001} & 0.0001 & 0.0001 & 0.0001 & 0.0001 & 0.0016 & 0.0016 & 0.0019 & 0.0026 & 0.0025 & 0.003 & 0.0052 & 0.0858 & 0.0108 & 0.0213 & 0.2354 \\ \hline
64 & \textbf{0.0057} & \textbf{0.0061} & 0.0065 & 0.0071 & 0.0087 & 0.0068 & 0.0081 & 0.0104 & 0.1103 & 0.0088 & 0.1489 & 0.0114 & 0.0472 & 0.0198 & 0.0156 & 0.1706 \\ \hline
65 & \textbf{0.0029} & \textbf{0.0029} & 0.003 & 0.0031 & 0.0035 & 0.0034 & 0.0037 & 0.0047 & 0.0322 & 0.0121 & 0.0794 & 0.0199 & 0.4521 & 0.0311 & 0.0813 & 10.2 \\ \hline
66 & \textbf{0.0029} & 0.0029 & 0.003 & 0.0031 & 0.0035 & 0.0034 & 0.0037 & 0.0047 & 0.0322 & 0.0121 & 0.0794 & 0.0199 & 0.4521 & 0.0311 & 0.0813 & 10.2 \\ \hline
67 & \textbf{0.0001} & 0.0002 & 0.0002 & 0.0002 & 0.0003 & 0.0002 & 0.0002 & 0.0003 & 0.0019 & 0.0007 & 0.0065 & 0.0012 & 0.0012 & 0.0009 & 0.001 & 0.0021 \\ \hline
\textbf{Avg} & 0.0029 & 0.003 & 0.0031 & 0.0034 & 0.004 & 0.0049 & 0.0055 & 0.0067 & 0.0406 & 0.0554 & 0.0766 & 0.1786 & 25.62 & 337.82 & 881.01 & 183786.02 \\ \hline
\end{tabular}
}}
\label{TableII}
\end{table}

\begin{table}[htbp]
\caption{Experiment II: RMSE per data set of the top 16 methods using 15\% of the samples using Lebesgue (L) and Riemann (R) sampling.}
\centering { \noindent\adjustbox{max width=\textwidth}{%
\begin{tabular}{|c|c|c|c|c|c|c|c|c|c|c|c|c|c|c|c|c|}
\hline
\textbf{N} & \textbf{L ZeChipC} & \textbf{L ZeLiC} & \textbf{L ZeChip} & \textbf{L ZeLi} & \textbf{L Zero} & \textbf{L PCHIP} & \textbf{L Linear} & \textbf{R PCHIP} & \textbf{R Quintic} & \textbf{R Cubic} & \textbf{L Nrst} & \textbf{R TP} & \textbf{R Mlt-qdc} & \textbf{R Lin} & \textbf{R Inv-mult} & \textbf{R Gauss} \\ \hline
1 & \textbf{0.0082} & 0.0096 & 0.0097 & 0.012 & 0.0159 & 0.0097 & 0.0123 & 0.0629 & 0.0607 & 0.069 & 0.0156 & 0.0745 & 0.0748 & 0.078 & 0.0833 & 0.0892 \\ \hline
2 & \textbf{0.0223} & 0.0241 & 0.0278 & 0.0322 & 0.0402 & 0.0285 & 0.0356 & 0.0564 & 0.0551 & 0.0625 & 0.0454 & 0.0675 & 0.0669 & 0.0701 & 0.0729 & 0.0772 \\ \hline
3 & 0.0102 & 0.0107 & 0.0119 & 0.0127 & 0.0141 & 0.0128 & 0.0144 & \textbf{0.0314} & 0.0317 & 0.0355 & 0.0179 & 0.0383 & 0.0377 & 0.0395 & 0.0418 & 0.0448 \\ \hline
4 & 0.0269 & 0.0274 & 0.0339 & 0.0397 & 0.0541 & 0.0337 & 0.0411 & \textbf{0.0172} & 0.0189 & 0.0191 & 0.0508 & 0.0194 & 0.0194 & 0.0186 & 0.0199 & 0.0213 \\ \hline
5 & 0.013 & 0.0133 & 0.0144 & 0.0158 & \textbf{0.0249} & 0.0144 & 0.017 & 0.0119 & 0.0128 & 0.0136 & 0.0235 & 0.0143 & 0.0142 & 0.0136 & 0.0153 & 0.0175 \\ \hline
6 & \textbf{0.0049} & 0.0047 & 0.0048 & 0.0046 & 0.0042 & 0.0075 & 0.0068 & 0.0061 & 0.0063 & 0.0064 & 0.0084 & 0.0064 & 0.0065 & 0.0063 & 0.0066 & 0.0068 \\ \hline
7 & \textbf{0.0027} & \textbf{0.0028} & \textbf{0.0029} & \textbf{0.0032} & \textbf{0.0052} & 0.003 & 0.0033 & 0.004 & 0.0041 & 0.0045 & 0.0048 & 0.0048 & 0.0048 & 0.0049 & 0.0054 & 0.0061 \\ \hline
8 & 0.001 & 0.001 & 0.001 & 0.001 & 0.001 & 0.0017 & 0.0017 & \textbf{0.0012} & 0.0014 & 0.0013 & 0.0019 & 0.0013 & 0.0013 & 0.0012 & 0.0013 & 0.0013 \\ \hline
9 & 0.0177 & 0.0189 & 0.0202 & 0.0221 & \textbf{0.0231} & 0.0224 & 0.0253 & 0.0159 & 0.0193 & 0.0188 & 0.0301 & 0.0187 & 0.0199 & 0.0181 & 0.0204 & 0.0227 \\ \hline
10 & \textbf{0.0011} & 0.0011 & 0.0011 & 0.0011 & 0.0007 & 0.0142 & 0.0161 & 0.0071 & 0.0075 & 0.0074 & 0.0196 & 0.0073 & 0.0075 & 0.007 & 0.0074 & 0.0075 \\ \hline
11 & \textbf{0.0013} & 0.0016 & 0.0017 & 0.0021 & 0.0032 & 0.0016 & 0.0021 & 0.0018 & 0.0019 & 0.0021 & 0.0029 & 0.0022 & 0.0022 & 0.0022 & 0.0026 & 0.0031 \\ \hline
12 & 0.0065 & 0.0066 & 0.0072 & 0.0077 & 0.008 & 0.0077 & 0.0086 & 0.0095 & \textbf{0.0084} & 0.0094 & 0.01 & 0.01 & 0.0099 & 0.0108 & 0.0107 & 0.0113 \\ \hline
13 & \textbf{0.0046} & 0.005 & 0.0053 & 0.0059 & 0.0054 & 0.0058 & 0.0065 & 0.0049 & 0.0043 & 0.0045 & 0.0071 & 0.0047 & 0.0046 & 0.0053 & 0.0049 & 0.0051 \\ \hline
14 & 0.0017 & \textbf{0.002} & 0.0018 & 0.002 & 0.0019 & 0.0019 & 0.0023 & 0.0086 & 0.0077 & 0.0091 & 0.0026 & 0.0098 & 0.0098 & 0.0105 & 0.0109 & 0.0115 \\ \hline
15 & \textbf{0.0117} & \textbf{0.0112} & \textbf{0.0118} & \textbf{0.0117} & 0.0121 & 0.013 & 0.0132 & 0.0184 & 0.0187 & 0.0189 & 0.016 & 0.019 & 0.0191 & 0.0185 & 0.0189 & 0.0186 \\ \hline
16 & \textbf{0.0002} & \textbf{0.0002} & \textbf{0.0002} & \textbf{0.0002} & 0.0003 & 0.0003 & 0.0004 & 0.0005 & 0.0005 & 0.0006 & 0.0005 & 0.0006 & 0.0006 & 0.0007 & 0.0007 & 0.0007 \\ \hline
17 & 0.0007 & 0.0007 & 0.0007 & 0.0007 & \textbf{0.001} & 0.002 & 0.0022 & 0.0035 & 0.0037 & 0.0037 & 0.0026 & 0.0037 & 0.0037 & 0.0035 & 0.0037 & 0.0039 \\ \hline
18 & \textbf{0.0251} & \textbf{0.0242} & \textbf{0.0252} & \textbf{0.0242} & \textbf{0.0116} & \textbf{0.0344} & \textbf{0.0312} & 0.0217 & 0.0229 & 0.0226 & 0.0374 & 0.0222 & 0.0227 & 0.0211 & 0.0224 & 0.0227 \\ \hline
19 & \textbf{0} & \textbf{0} & 0 & 0 & \textbf{0} & 0 & 0 & 0.0003 & 0.0003 & 0.0003 & 0.0001 & 0.0003 & 0.0003 & 0.0003 & 0.0003 & 0.0003 \\ \hline
20 & 0.0018 & 0.0018 & 0.0019 & 0.0019 & \textbf{0.0018} & 0.0021 & 0.0021 & 0.0029 & 0.0031 & 0.003 & 0.0024 & 0.003 & 0.0031 & 0.0029 & 0.003 & 0.0031 \\ \hline
21 & 0.0167 & 0.0161 & 0.0171 & 0.0168 & \textbf{0.0151} & 0.0252 & 0.0269 & 0.021 & 0.0211 & 0.021 & 0.031 & 0.0212 & 0.0209 & 0.0216 & 0.0211 & 0.0213 \\ \hline
22 & \textbf{0.0013} & \textbf{0.0013} & 0.0014 & 0.0014 & 0.0012 & 0.0017 & 0.0016 & 0.0023 & 0.0024 & 0.0024 & 0.0019 & 0.0024 & 0.0025 & 0.0024 & 0.0026 & 0.0027 \\ \hline
23 & 0.0009 & 0.0009 & 0.001 & 0.0011 & 0.0019 & 0.001 & 0.0012 & \textbf{0.002} & \textbf{0.002} & \textbf{0.0023} & 0.0017 & \textbf{0.0024} & \textbf{0.0024} & 0.0025 & \textbf{0.0028} & \textbf{0.0031} \\ \hline
24 & 0.0006 & 0.0006 & 0.0007 & 0.0007 & 0.0006 & 0.0008 & 0.0008 & \textbf{0.0005} & \textbf{0.0005} & \textbf{0.0005} & 0.0009 & \textbf{0.0005} & \textbf{0.0005} & 0.0006 & \textbf{0.0005} & \textbf{0.0005} \\ \hline
25 & \textbf{0.0006} & 0.0007 & 0.0007 & 0.0008 & \textbf{0.0007} & 0.0008 & 0.0009 & 0.0005 & 0.0005 & 0.0005 & 0.001 & 0.0005 & 0.0005 & 0.0006 & 0.0005 & 0.0005 \\ \hline
26 & \textbf{0.007} & \textbf{0.0074} & 0.0074 & 0.0079 & 0.007 & \textbf{0.0088} & 0.0105 & 0.0088 & 0.01 & 0.01 & 0.0122 & 0.01 & 0.0102 & 0.0097 & 0.0102 & 0.0105 \\ \hline
27 & \textbf{0} & \textbf{0} & \textbf{0.0001} & 0.0001 & 0.0002 & 0 & 0.0001 & 0.0004 & 0.0004 & 0.0005 & 0.0001 & 0.0005 & 0.0005 & 0.0005 & 0.0006 & 0.0006 \\ \hline
28 & 0.0002 & 0.0002 & 0.0002 & 0.0003 & 0.0004 & 0.0003 & 0.0003 & \textbf{0.0019} & 0.0026 & 0.0027 & 0.0004 & 0.0029 & 0.0028 & 0.0024 & 0.003 & 0.0031 \\ \hline
29 & \textbf{0.0036} & \textbf{0.0037} & \textbf{0.0041} & \textbf{0.0046} & \textbf{0.0075} & \textbf{0.0042} & \textbf{0.0047} & 0.0034 & 0.0036 & 0.0039 & 0.0067 & 0.0041 & 0.004 & 0.004 & 0.004 & 0.0043 \\ \hline
30 & \textbf{0.0002} & 0.0002 & 0.0002 & 0.0002 & 0.0002 & 0.0002 & 0.0002 & 0.001 & 0.001 & 0.0011 & 0.0003 & 0.0012 & 0.0012 & 0.0013 & 0.0014 & 0.0015 \\ \hline
31 & 0.0003 & 0.0004 & \textbf{0.0004} & 0.0004 & 0.0004 & 0.0004 & 0.0005 & 0.001 & 0.001 & 0.0011 & 0.0006 & 0.0012 & 0.0012 & 0.0013 & 0.0013 & 0.0013 \\ \hline
32 & \textbf{0.0088} & \textbf{0.009} & \textbf{0.0087} & \textbf{0.0089} & \textbf{0.0088} & 0.0093 & 0.0092 & 0.0109 & 0.0154 & 0.0117 & 0.0111 & 0.0116 & 0.0114 & 0.012 & 0.0123 & 0.0129 \\ \hline
33 & \textbf{0} & \textbf{0} & \textbf{0} & \textbf{0} & \textbf{0} & \textbf{0.0015} & \textbf{0.0017} & 0.0032 & 0.0035 & 0.0034 & 0.0019 & 0.0034 & 0.0035 & 0.0032 & 0.0034 & 0.0035 \\ \hline
34 & \textbf{0.0002} & 0.0002 & 0.0002 & 0.0002 & 0.0002 & 0.0002 & 0.0002 & 0.0005 & 0.0005 & 0.0006 & 0.0003 & 0.0006 & 0.0006 & 0.0006 & 0.0007 & 0.0007 \\ \hline
35 & 0.0027 & \textbf{0.0028} & 0.0029 & \textbf{0.0032} & 0.0035 & 0.0032 & 0.0037 & 0.006 & 0.0061 & 0.0068 & 0.0046 & 0.0074 & 0.0073 & 0.0076 & 0.0081 & 0.0088 \\ \hline
36 & \textbf{0.0007} & 0.0006 & 0.0007 & 0.0006 & 0.0008 & 0.0008 & 0.0007 & 0.0036 & 0.0036 & 0.0037 & 0.0009 & 0.0038 & 0.0038 & 0.0039 & 0.004 & 0.0042 \\ \hline
37 & 0.006 & 0.0066 & 0.0063 & 0.007 & 0.0067 & 0.0066 & 0.0078 & 0.0088 & \textbf{0.0072} & 0.0085 & 0.009 & 0.0092 & 0.0089 & 0.01 & 0.0097 & 0.0103 \\ \hline
38 & \textbf{0.0043} & 0.0047 & 0.0045 & 0.0049 & 0.0047 & 0.0048 & 0.0056 & 0.0041 & 0.0034 & 0.0038 & 0.0063 & 0.004 & 0.0038 & 0.0045 & 0.0041 & 0.0043 \\ \hline
39 & \textbf{0.0019} & \textbf{0.0022} & \textbf{0.002} & 0.0023 & 0.002 & 0.0021 & 0.0026 & 0.0084 & 0.0075 & 0.0088 & 0.0029 & 0.0095 & 0.0096 & 0.0102 & 0.0106 & 0.0112 \\ \hline
40 & \textbf{0.0017} & 0.0017 & 0.0017 & 0.0018 & 0.002 & 0.0047 & 0.0045 & 0.0036 & 0.0037 & 0.0038 & 0.0054 & 0.0038 & 0.0039 & 0.0038 & 0.004 & 0.0042 \\ \hline
41 & \textbf{0.0018} & 0.0019 & 0.002 & 0.0022 & 0.003 & 0.0021 & 0.0023 & 0.0021 & 0.0021 & 0.0023 & 0.0029 & 0.0024 & 0.0024 & 0.0025 & 0.0027 & 0.0029 \\ \hline
42 & 0.0087 & 0.0094 & 0.009 & 0.0103 & 0.0112 & 0.0099 & 0.0122 & 0.0221 & \textbf{0.0226} & 0.0249 & 0.015 & 0.0268 & 0.0263 & 0.0282 & 0.0287 & 0.0299 \\ \hline
43 & \textbf{0.0015} & \textbf{0.0016} & \textbf{0.0017} & \textbf{0.0019} & \textbf{0.0018} & 0.0019 & 0.0021 & 0.0016 & 0.0014 & 0.0015 & 0.0023 & 0.0016 & 0.0015 & 0.0017 & 0.0016 & 0.0017 \\ \hline
44 & 0.0009 & 0.0009 & 0.0009 & 0.0009 & 0.0009 & 0.0011 & 0.0012 & 0.0012 & \textbf{0.0013} & 0.0012 & 0.0014 & 0.0012 & 0.0013 & 0.0012 & 0.0013 & 0.0013 \\ \hline
45 & \textbf{0.0094} & 0.0096 & 0.0104 & 0.0108 & 0.011 & 0.0107 & 0.0113 & 0.008 & 0.0078 & 0.0084 & 0.0138 & 0.0089 & 0.009 & 0.0093 & 0.0098 & 0.0107 \\ \hline
46 & 0.0057 & 0.0065 & 0.0059 & 0.0068 & 0.0068 & 0.0061 & 0.0073 & 0.0086 & \textbf{0.0068} & 0.0083 & 0.0085 & 0.0089 & 0.0086 & 0.0097 & 0.0094 & 0.01 \\ \hline
47 & \textbf{0.0045} & 0.0051 & 0.0047 & 0.0053 & 0.0053 & 0.0049 & 0.0057 & 0.0046 & 0.0035 & 0.0042 & 0.0065 & 0.0045 & 0.0043 & 0.0051 & 0.0046 & 0.0049 \\ \hline
48 & 0.002 & 0.0022 & 0.0022 & 0.0026 & \textbf{0.0029} & 0.0022 & 0.0026 & 0.008 & 0.0069 & 0.0082 & 0.0032 & 0.0089 & 0.0089 & 0.0096 & 0.0098 & 0.0104 \\ \hline
49 & 0.0024 & 0.0024 & 0.0023 & 0.0024 & \textbf{0.0021} & 0.0133 & 0.0139 & 0.0084 & 0.0089 & 0.0088 & 0.0164 & 0.0087 & 0.0088 & 0.0084 & 0.0088 & 0.0089 \\ \hline
50 & 0.0006 & 0.0006 & 0.0006 & 0.0006 & \textbf{0.0004} & 0.0069 & 0.0094 & 0.0042 & 0.0045 & 0.0045 & 0.0114 & 0.0044 & 0.0045 & 0.0043 & 0.0045 & 0.0046 \\ \hline
51 & \textbf{0.0639} & \textbf{0.06} & \textbf{0.0644} & 0.0606 & 0.0486 & \textbf{0.0672} & 0.0618 & 0.067 & 0.0707 & 0.0696 & 0.0723 & 0.0684 & 0.07 & 0.0648 & 0.0692 & 0.07 \\ \hline
52 & \textbf{0.0003} & \textbf{0.0003} & \textbf{0.0003} & \textbf{0.0004} & \textbf{0.0006} & 0.0003 & 0.0004 & 0.0004 & 0.0004 & 0.0005 & 0.0006 & 0.0005 & 0.0005 & 0.0005 & 0.0006 & 0.0007 \\ \hline
53 & \textbf{0} & \textbf{0} & \textbf{0} & \textbf{0} & \textbf{0} & \textbf{0.013} & \textbf{0.0137} & 0.0046 & 0.0048 & 0.0048 & \textbf{0.0167} & 0.0047 & 0.0048 & 0.0046 & 0.0048 & 0.0048 \\ \hline
54 & \textbf{0} & 0 & 0 & 0 & 0 & 0 & 0 & 0.0001 & 0.0001 & 0.0001 & 0 & 0.0001 & 0.0001 & 0.0001 & 0.0001 & 0.0001 \\ \hline
55 & \textbf{0.0009} & 0.001 & 0.001 & 0.0011 & 0.0012 & 0.0011 & 0.0012 & 0.0019 & 0.0019 & 0.0021 & 0.0015 & 0.0022 & 0.0022 & 0.0023 & 0.0024 & 0.0026 \\ \hline
56 & \textbf{0.0008} & \textbf{0.0009} & \textbf{0.0009} & \textbf{0.001} & 0.0013 & \textbf{0.0009} & 0.0011 & 0.002 & 0.0019 & 0.0022 & 0.0013 & 0.0023 & 0.0024 & 0.0025 & 0.0027 & 0.0029 \\ \hline
57 & \textbf{0.0002} & \textbf{0.0002} & 0.0002 & 0.0002 & 0.0005 & 0.0002 & 0.0003 & 0.0015 & 0.0014 & 0.0016 & 0.0004 & 0.0018 & 0.0018 & 0.0019 & 0.002 & 0.0021 \\ \hline
58 & \textbf{0.0058} & \textbf{0.0058} & 0.006 & 0.0061 & 0.0063 & 0.007 & 0.0077 & 0.0064 & 0.0065 & 0.0066 & 0.0094 & 0.0068 & 0.0067 & 0.0071 & 0.0069 & 0.0071 \\ \hline
59 & 0.0067 & 0.0067 & 0.0069 & 0.0071 & \textbf{0.0076} & 0.0089 & 0.0093 & 0.0075 & 0.008 & 0.0083 & 0.0116 & 0.0085 & 0.0085 & 0.0085 & 0.0088 & 0.0091 \\ \hline
60 & \textbf{0.0015} & \textbf{0.0014} & 0.0014 & 0.0014 & 0.0013 & 0.0018 & 0.002 & 0.0111 & 0.0112 & 0.012 & 0.0024 & 0.0126 & 0.0127 & 0.013 & 0.0136 & 0.0143 \\ \hline
61 & \textbf{0.0013} & \textbf{0.0013} & \textbf{0.0015} & \textbf{0.0016} & \textbf{0.0018} & 0.0028 & 0.0029 & 0.0023 & 0.0024 & 0.0025 & 0.0035 & 0.0025 & 0.0025 & 0.0026 & 0.0027 & 0.0029 \\ \hline
62 & \textbf{0.0001} & \textbf{0.0001} & \textbf{0.0001} & \textbf{0.0001} & \textbf{0.0001} & 0.0002 & 0.0002 & 0.0002 & 0.0002 & 0.0003 & 0.0003 & 0.0003 & 0.0003 & 0.0003 & 0.0003 & 0.0003 \\ \hline
63 & \textbf{0} & 0 & 0 & 0 & 0 & 0.0012 & 0.0012 & 0.0007 & 0.0008 & 0.0008 & 0.0015 & 0.0008 & 0.0008 & 0.0007 & 0.0008 & 0.0008 \\ \hline
64 & \textbf{0.01} & 0.0118 & 0.0104 & 0.0122 & 0.0109 & 0.012 & 0.0158 & 0.0109 & 0.0124 & 0.0121 & 0.018 & 0.0117 & 0.0123 & 0.0108 & 0.0118 & 0.012 \\ \hline
65 & \textbf{0.0025} & 0.0026 & 0.0026 & 0.0028 & 0.0031 & 0.003 & 0.0032 & 0.0028 & 0.0029 & 0.003 & 0.0041 & 0.0031 & 0.0031 & 0.0031 & 0.0033 & 0.0035 \\ \hline
66 & \textbf{0.0025} & \textbf{0.0026} & \textbf{0.0026} & \textbf{0.0028} & 0.0031 & \textbf{0.003} & \textbf{0.0032} & \textbf{0.0028} & \textbf{0.0029} & \textbf{0.003} & 0.0041 & \textbf{0.0031} & \textbf{0.0031} & \textbf{0.0031} & \textbf{0.0033} & 0.0035 \\ \hline
67 & \textbf{0.0002} & 0.0002 & 0.0002 & 0.0002 & 0.0003 & 0.0002 & 0.0002 & 0.0002 & 0.0002 & 0.0002 & 0.0003 & 0.0002 & 0.0002 & 0.0002 & 0.0002 & 0.0003 \\ \hline
Avg & \textbf{0.0053} & \textbf{0.0054} & \textbf{0.0057} & \textbf{0.0061} & \textbf{0.0063} & \textbf{0.0071} & \textbf{0.0077} & \textbf{0.0085} & \textbf{0.0087} & \textbf{0.0091} & \textbf{0.0094} & \textbf{0.0095} & \textbf{0.0095} & \textbf{0.0096} & \textbf{0.0100} & \textbf{0.0105} \\ \hline
\end{tabular}

}}
\label{TableIII}
\end{table}

\begin{table}[htbp]
\caption{Experiment II: RMSE per data set of the last 16 methods using 15\% of the samples using Lebesgue (L) and Riemann (R) sampling.}
\centering { \noindent\adjustbox{max width=\textwidth}{%
\begin{tabular}{|c|c|c|c|c|c|c|c|c|c|c|c|c|c|c|c|c|}
\hline
\textbf{N} & \textbf{R Nrst} & \textbf{R ZeLi} & \textbf{R ZeChip} & \textbf{R ZeLiC} & \textbf{R ZeChipC} & \textbf{R Zero} & \textbf{R Shannon} & \textbf{L T-P} & \textbf{L Shannon} & \textbf{R Lasso} & \textbf{L Lasso} & \textbf{L Cubic} & \textbf{L Inv mlt} & \textbf{L Quintic} & \textbf{L Mltqdc} & \textbf{L Gssn} \\ \hline
1 & 0.1026 & 0.1472 & 0.1472 & 0.1472 & 0.1472 & 0.1475 & 0.1419 & 0.1287 & 0.1848 & 0.1613 & 0.1613 & 0.241 & 0.3153 & 0.3727 & 0.3499 & 0.4319 \\ \hline
2 & 0.0931 & 0.1328 & 0.1328 & 0.1331 & 0.1334 & 0.1349 & 0.1374 & 0.0913 & 0.2531 & 0.6052 & 0.6052 & 0.1459 & 0.1881 & 0.1416 & 0.1677 & 0.342 \\ \hline
3 & 0.0529 & 0.0763 & 0.0763 & 0.0763 & 0.0764 & 0.0767 & 0.0969 & 0.1238 & 0.2195 & 0.4262 & 0.4262 & 0.2496 & 0.1858 & 0.6928 & 0.3786 & 2236.4 \\ \hline
4 & 0.0433 & 0.0634 & 0.0641 & 0.065 & 0.067 & 0.0737 & 0.0808 & 0.0337 & 0.1861 & 0.5401 & 0.5401 & 0.038 & 0.0477 & 0.047 & 0.0377 & 0.1169 \\ \hline
5 & 0.0254 & 0.038 & 0.0381 & 0.0384 & 0.0387 & 0.0408 & 0.044 & 0.0239 & 0.1687 & 0.5603 & 0.5603 & 0.0368 & 0.057 & 0.0455 & 0.0367 & 0.1941 \\ \hline
6 & 0.0074 & 0.0089 & 0.009 & 0.0092 & 0.0094 & 0.0091 & 0.0087 & 0.0152 & 0.0123 & 0.0209 & 0.0209 & 0.0313 & 0.0248 & 0.1701 & 0.0468 & 0.0703 \\ \hline
7 & 0.0074 & 0.0109 & 0.0109 & 0.0109 & 0.011 & 0.0114 & 0.0118 & 0.0049 & 0.0354 & 0.1527 & 0.1527 & 0.008 & 0.0157 & 0.0124 & 0.0125 & 0.0473 \\ \hline
8 & 0.0014 & 0.0015 & 0.0015 & 0.0016 & 0.0016 & 0.0016 & 0.0018 & 0.0042 & 0.0028 & 0.0025 & 0.0025 & 0.0089 & 0.0082 & 0.0555 & 0.0247 & 0.0515 \\ \hline
9 & 0.0331 & 0.0544 & 0.0545 & 0.0561 & 0.057 & 0.0561 & 0.0745 & 0.0277 & 0.1427 & 0.3963 & 0.3963 & 0.0797 & 0.0496 & 0.2464 & 0.0811 & 7.181 \\ \hline
10 & 0.0084 & 0.0099 & 0.0099 & 0.01 & 0.01 & 0.01 & 0.0105 & 0.3398 & 0.0428 & 0.0222 & 0.0222 & 1.784 & 2908.7 & 677 & 11160 & 1669510.8 \\ \hline
11 & 0.0038 & 0.0055 & 0.0055 & 0.0055 & 0.0055 & 0.006 & 0.0061 & 0.0029 & 0.0139 & 0.0495 & 0.0495 & 0.0043 & 0.0073 & 0.0041 & 0.0054 & 0.0106 \\ \hline
12 & 0.0129 & 0.0159 & 0.0158 & 0.0166 & 0.0168 & 0.0179 & 0.0174 & 0.0103 & 0.0226 & 0.035 & 0.035 & 0.0211 & 0.0166 & 0.1473 & 0.0205 & 0.0303 \\ \hline
13 & 0.0065 & 0.0077 & 0.0078 & 0.0084 & 0.0087 & 0.0095 & 0.0092 & 0.006 & 0.0122 & 0.0248 & 0.0248 & 0.0071 & 0.007 & 0.02 & 0.0075 & 0.0115 \\ \hline
14 & 0.0127 & 0.0175 & 0.0175 & 0.0175 & 0.0175 & 0.0175 & 0.0176 & 0.0142 & 0.0195 & 0.0139 & 0.0139 & 0.0494 & 0.0117 & 0.3679 & 0.019 & 0.0306 \\ \hline
15 & 0.0214 & 0.024 & 0.0241 & 0.0246 & 0.025 & 0.0249 & 0.026 & 0.0235 & 0.0622 & 0.091 & 0.091 & 0.0803 & 0.0704 & 1.187 & 0.1414 & 2.125 \\ \hline
16 & 0.0008 & 0.0012 & 0.0012 & 0.0012 & 0.0012 & 0.0012 & 0.001 & 0.0005 & 0.0021 & 0.0052 & 0.0052 & 0.0014 & 0.003 & 0.0098 & 0.0138 & 0.4805 \\ \hline
17 & 0.0039 & 0.0051 & 0.0051 & 0.0051 & 0.0052 & 0.0051 & 0.0051 & 0.0098 & 0.0112 & 0.0297 & 0.0297 & 0.0277 & 0.0869 & 0.3067 & 0.455 & 36.4 \\ \hline
18 & 0.0233 & 0.0215 & 0.0222 & 0.0232 & 0.0242 & 0.0235 & 0.0243 & 0.0972 & 0.0362 & 0.0208 & 0.0208 & 0.5384 & 0.6116 & 363 & 1.787 & 300.3 \\ \hline
19 & 0.0003 & 0.0003 & 0.0003 & 0.0003 & 0.0003 & 0.0003 & 0.0004 & 0.0042 & 0.001 & 0.0003 & 0.0003 & 0.0123 & 0.2195 & 1.502 & 1.408 & 30.6 \\ \hline
20 & 0.0032 & 0.0042 & 0.0042 & 0.0044 & 0.0045 & 0.0043 & 0.0034 & 0.0048 & 0.0051 & 0.0089 & 0.0089 & 0.0211 & 0.0038 & 0.3991 & 0.0078 & 0.0119 \\ \hline
21 & 0.0302 & 0.0446 & 0.0447 & 0.0455 & 0.046 & 0.0456 & 0.0566 & 0.0466 & 0.0938 & 0.1725 & 0.1725 & 0.1006 & 0.0818 & 0.5599 & 0.214 & 8.027 \\ \hline
22 & 0.0028 & 0.0036 & 0.0036 & 0.0036 & 0.0037 & 0.0037 & 0.0039 & 0.0254 & 0.0048 & 0.0059 & 0.0059 & 1.005 & 0.0334 & 12127.4 & 0.0967 & 0.1279 \\ \hline
23 & 0.0036 & 0.0052 & 0.0052 & 0.0052 & 0.0052 & 0.0054 & 0.0053 & 0.003 & 0.0104 & 0.0393 & 0.0393 & 0.0047 & 0.0073 & 0.0048 & 0.0059 & 0.0139 \\ \hline
24 & 0.0008 & 0.0013 & 0.0013 & 0.0013 & 0.0013 & 0.0013 & 0.0015 & 0.0012 & 0.0021 & 0.0041 & 0.0041 & 0.0017 & 0.0009 & 0.0049 & 0.0011 & 0.0058 \\ \hline
25 & 0.0009 & 0.0014 & 0.0014 & 0.0014 & 0.0014 & 0.0014 & 0.0016 & 0.0009 & 0.0022 & 0.0046 & 0.0046 & 0.0013 & 0.0009 & 0.0039 & 0.0011 & 0.0023 \\ \hline
26 & 0.0138 & 0.021 & 0.021 & 0.0212 & 0.0214 & 0.0213 & 0.026 & 0.0632 & 0.0582 & 0.0509 & 0.0509 & 0.1345 & 0.4474 & 0.3428 & 2.561 & 135.6 \\ \hline
27 & 0.0007 & 0.001 & 0.001 & 0.001 & 0.001 & 0.001 & 0.0009 & 0.0044 & 0.0015 & 0.0178 & 0.0178 & 0.0092 & 0.0023 & 0.0091 & 0.0059 & 0.0044 \\ \hline
28 & 0.0028 & 0.0028 & 0.0028 & 0.0028 & 0.0028 & 0.0028 & 0.0054 & 0.002 & 0.0096 & 0.0575 & 0.0575 & 0.0023 & 0.0203 & 0.0106 & 0.085 & 147.6 \\ \hline
29 & 0.0071 & 0.0104 & 0.0105 & 0.0105 & 0.0106 & 0.0115 & 0.012 & 0.0049 & 0.0457 & 0.1642 & 0.1642 & 0.0059 & 0.0106 & 0.0066 & 0.0064 & 0.039 \\ \hline
30 & 0.0017 & 0.0023 & 0.0023 & 0.0023 & 0.0023 & 0.0023 & 0.0023 & 0.0009 & 0.0098 & 0.0196 & 0.0196 & 0.0009 & 0.0229 & 99.8 & 0.0536 & 151643.5 \\ \hline
31 & 0.0017 & 0.0025 & 0.0025 & 0.0025 & 0.0025 & 0.0025 & 0.0022 & 0.0086 & 0.0056 & 0.0031 & 0.0031 & 0.0289 & 0.0077 & 0.4018 & 0.0272 & 183.4 \\ \hline
32 & 0.0138 & 0.0126 & 0.0128 & 0.0136 & 0.0139 & 0.0172 & 0.0123 & 0.0127 & 0.013 & 0.0208 & 0.0208 & 0.4598 & 0.0183 & 178 & 0.0271 & 0.022 \\ \hline
33 & 0.0038 & 0.0046 & 0.0046 & 0.0046 & 0.0046 & 0.0046 & 0.0049 & 0.1053 & 0.0157 & 0.0058 & 0.0058 & 0.4462 & 6227.8 & 145 & 391795.3 & 6219488.3 \\ \hline
34 & 0.0008 & 0.0012 & 0.0012 & 0.0012 & 0.0012 & 0.0012 & 0.0012 & 0.0035 & 0.003 & 0.005 & 0.005 & 0.0069 & 0.0072 & 0.008 & 0.0222 & 202.6 \\ \hline
35 & 0.011 & 0.0164 & 0.0164 & 0.0164 & 0.0164 & 0.0164 & 0.0221 & 0.0574 & 0.1053 & 0.1191 & 0.1191 & 0.1304 & 0.0408 & 0.3145 & 0.0776 & 12374.5 \\ \hline
36 & 0.0046 & 0.0059 & 0.0059 & 0.0059 & 0.0059 & 0.0059 & 0.0065 & 0.0248 & 0.0111 & 0.0066 & 0.0066 & 0.1927 & 0.0189 & 34.2 & 0.0871 & 5.719 \\ \hline
37 & 0.0117 & 0.0146 & 0.0146 & 0.0151 & 0.0153 & 0.0158 & 0.0156 & 0.0082 & 0.0236 & 0.0346 & 0.0346 & 0.0203 & 0.012 & 0.1448 & 0.0125 & 0.0255 \\ \hline
38 & 0.0057 & 0.0074 & 0.0074 & 0.0079 & 0.0081 & 0.0083 & 0.0078 & 0.0047 & 0.0125 & 0.0246 & 0.0246 & 0.0071 & 0.0055 & 0.0336 & 0.0052 & 0.0094 \\ \hline
39 & 0.0124 & 0.0169 & 0.017 & 0.017 & 0.017 & 0.017 & 0.0171 & 0.0123 & 0.0193 & 0.0142 & 0.0142 & 0.0414 & 0.0127 & 0.3326 & 0.0182 & 0.0347 \\ \hline
40 & 0.0047 & 0.0061 & 0.0061 & 0.0061 & 0.0062 & 0.0062 & 0.0057 & 0.0157 & 0.0097 & 0.0181 & 0.0181 & 0.0531 & 0.0357 & 1.507 & 0.0803 & 0.073 \\ \hline
41 & 0.0039 & 0.0057 & 0.0057 & 0.0057 & 0.0058 & 0.006 & 0.0063 & 0.004 & 0.0146 & 0.0378 & 0.0378 & 0.0065 & 0.007 & 0.0084 & 0.0079 & 0.014 \\ \hline
42 & 0.04 & 0.0605 & 0.0604 & 0.0605 & 0.0606 & 0.0605 & 0.0732 & 0.1169 & 0.2198 & 0.1931 & 0.1931 & 0.38 & 0.162 & 1.099 & 1.863 & 5321893.7 \\ \hline
43 & 0.0022 & 0.0025 & 0.0026 & 0.0028 & 0.0029 & 0.0031 & 0.003 & 0.002 & 0.004 & 0.0082 & 0.0082 & 0.0023 & 0.0023 & 0.0066 & 0.0025 & 0.0038 \\ \hline
44 & 0.0015 & 0.0019 & 0.0019 & 0.0019 & 0.002 & 0.0019 & 0.0019 & 0.0033 & 0.0043 & 0.0097 & 0.0097 & 0.0166 & 0.0126 & 2.373 & 0.0236 & 742.7 \\ \hline
45 & 0.0146 & 0.022 & 0.0221 & 0.0222 & 0.0224 & 0.0233 & 0.0234 & 0.0114 & 0.0417 & 0.0604 & 0.0604 & 0.0133 & 0.0178 & 0.0307 & 0.0247 & 0.0423 \\ \hline
46 & 0.0116 & 0.0148 & 0.0149 & 0.0155 & 0.0158 & 0.0159 & 0.0148 & 0.0094 & 0.0257 & 0.0308 & 0.0308 & 0.0287 & 0.0112 & 0.2332 & 0.0125 & 0.0236 \\ \hline
47 & 0.0065 & 0.0081 & 0.0082 & 0.0088 & 0.0092 & 0.0094 & 0.0082 & 0.0052 & 0.0167 & 0.0235 & 0.0235 & 0.0115 & 0.0062 & 0.0806 & 0.0065 & 0.012 \\ \hline
48 & 0.0115 & 0.0156 & 0.0156 & 0.0156 & 0.0157 & 0.0158 & 0.0155 & 0.0105 & 0.0199 & 0.0151 & 0.0151 & 0.0364 & 0.0123 & 0.3198 & 0.0148 & 0.0262 \\ \hline
49 & 0.0101 & 0.0126 & 0.0126 & 0.0128 & 0.0128 & 0.0126 & 0.0133 & 0.0274 & 0.0254 & 0.0239 & 0.0239 & 0.0787 & 0.093 & 0.4242 & 0.2576 & 54 \\ \hline
50 & 0.0053 & 0.0066 & 0.0066 & 0.0067 & 0.0067 & 0.0066 & 0.007 & 0.309 & 0.0295 & 0.0201 & 0.0201 & 1.031 & 106.3 & 87 & 4123.5 & 6157899.9 \\ \hline
51 & 0.072 & 0.0674 & 0.0696 & 0.0736 & 0.0774 & 0.0719 & 0.0744 & 0.2912 & 0.0781 & 0.1102 & 0.1102 & 3.958 & 37.2 & 5330 & 389.3 & 41060.8 \\ \hline
52 & 0.0008 & 0.0013 & 0.0013 & 0.0013 & 0.0013 & 0.0013 & 0.0013 & 0.0008 & 0.0064 & 0.0167 & 0.0167 & 0.002 & 0.0022 & 0.0047 & 0.0015 & 0.0064 \\ \hline
53 & 0.0049 & 0.005 & 0.005 & 0.005 & 0.005 & 0.005 & 0.005 & 0.1015 & 0.0209 & 0.0039 & 0.0039 & 0.8011 & 9554.9 & 124.3 & 15063.9 & 116244 \\ \hline
54 & 0.0001 & 0.0002 & 0.0002 & 0.0002 & 0.0002 & 0.0002 & 0.0002 & 0.0002 & 0.0007 & 0.0018 & 0.0018 & 0.0006 & 0.0003 & 0.0011 & 0.0004 & 0.9184 \\ \hline
55 & 0.0031 & 0.0045 & 0.0045 & 0.0046 & 0.0046 & 0.0046 & 0.0048 & 0.0063 & 0.01 & 0.016 & 0.016 & 0.0107 & 0.0084 & 0.0111 & 0.0111 & 0.1093 \\ \hline
56 & 0.0034 & 0.0051 & 0.0051 & 0.0051 & 0.0051 & 0.0051 & 0.0048 & 0.0029 & 0.0063 & 0.0082 & 0.0082 & 0.0055 & 0.0072 & 0.0135 & 0.0104 & 0.0142 \\ \hline
57 & 0.0024 & 0.0034 & 0.0034 & 0.0034 & 0.0034 & 0.0034 & 0.0027 & 0.0059 & 0.0046 & 0.0162 & 0.0162 & 0.0118 & 0.0067 & 0.0363 & 0.0135 & 0.0535 \\ \hline
58 & 0.0104 & 0.0156 & 0.0156 & 0.0158 & 0.0159 & 0.0159 & 0.0169 & 0.0125 & 0.0375 & 0.0632 & 0.0632 & 0.021 & 0.0243 & 0.0785 & 0.0652 & 0.4154 \\ \hline
59 & 0.0129 & 0.019 & 0.019 & 0.0192 & 0.0193 & 0.0194 & 0.0215 & 0.0457 & 0.0629 & 0.0978 & 0.0978 & 0.09 & 0.0393 & 0.1938 & 0.1052 & 872.5 \\ \hline
60 & 0.0164 & 0.0232 & 0.0232 & 0.0232 & 0.0232 & 0.0232 & 0.018 & 0.1616 & 0.0341 & 0.078 & 0.078 & 0.4371 & 0.4643 & 1.885 & 3.87 & 1106.6 \\ \hline
61 & 0.0034 & 0.0049 & 0.0049 & 0.0049 & 0.005 & 0.005 & 0.0049 & 0.0098 & 0.0085 & 0.0221 & 0.0221 & 0.0136 & 0.0094 & 0.0417 & 0.0149 & 0.0162 \\ \hline
62 & 0.0004 & 0.0005 & 0.0005 & 0.0005 & 0.0005 & 0.0005 & 0.0005 & 0.0034 & 0.0015 & 0.003 & 0.003 & 0.0072 & 0.0026 & 0.0105 & 0.0068 & 11.2 \\ \hline
63 & 0.0009 & 0.0012 & 0.0012 & 0.0012 & 0.0012 & 0.0012 & 0.0011 & 0.002 & 0.0024 & 0.003 & 0.003 & 0.0035 & 0.0112 & 0.0256 & 0.0334 & 0.3011 \\ \hline
64 & 0.0164 & 0.0278 & 0.0278 & 0.0284 & 0.0286 & 0.0282 & 0.0367 & 0.0278 & 0.1123 & 0.1489 & 0.1489 & 0.0555 & 0.0336 & 0.4097 & 0.0606 & 0.4505 \\ \hline
65 & 0.0046 & 0.0066 & 0.0066 & 0.0067 & 0.0067 & 0.0068 & 0.0074 & 0.0104 & 0.0306 & 0.0794 & 0.0794 & 0.0182 & 0.0226 & 0.1942 & 0.0532 & 9.207 \\ \hline
66 & 0.0046 & 0.0066 & 0.0066 & 0.0067 & 0.0067 & 0.0068 & 0.0074 & 0.0104 & 0.0306 & 0.0794 & 0.0794 & 0.0182 & 0.0226 & 0.1942 & 0.0532 & 9.207 \\ \hline
67 & 0.0004 & 0.0005 & 0.0005 & 0.0005 & 0.0005 & 0.0006 & 0.0006 & 0.0007 & 0.0019 & 0.0065 & 0.0065 & 0.0012 & 0.0009 & 0.0012 & 0.0011 & 0.0022 \\ \hline
Avg & \textbf{0.0129} & \textbf{0.0175} & \textbf{0.0176} & \textbf{0.0178} & \textbf{0.0180} & \textbf{0.0182} & \textbf{0.0195} & \textbf{0.0383} & \textbf{0.0409} & \textbf{0.0766} & \textbf{0.0766} & \textbf{0.1955} & \textbf{281.17} & \textbf{286.31} & \textbf{6306.67} & \textbf{293973.46} \\ \hline
\end{tabular}
}}
\label{TableIV}
\end{table}

\begin{algorithm}
\caption{ZeLiC algorithm using mathematical nomenclature}\label{alg:zelicMat}
\textbf{Input: }
$\left(x_j,y_j\right) \forall j=i-1,\,i,\,i+1$, $t$ the threshold, the parameters seen before
\begin{algorithmic}[2]
\State $Condition\, 1 \gets subsequent\_min\_distance < x_{i+1} - x_{i} < subsequent\_max\_distance$
\State $Condition\, 2 \gets x_{i}-x_{i-1} > previous\_distance$
\State $Condition\, 3 \gets sign\left(y_{i}-y_{i-1}\right) \neq sign\left(y_{i+1} - y_{i}\right) $ \Comment $sign$ returns the sign of the expression
\State $a \gets \frac{y_{i+1}-y_i}{x_{i+1}-x_i}$
\State $b \gets y_i - ax_i$
\State $Limit\,Condition \gets x_{i,n}>\frac{t}{\left|a\right|}+x_i$\eqref{limit_cond}
\If{$Condition\,1$ AND $Condition\,2$ AND $Condition\,3$}
    \State $sign\_before \gets sign\left(y_{i}-y_{i-1}\right)$
    \State $x_{i+\frac{1}{2}} \gets \frac{x_{i+1} + x_{i}}{2}$
    \If{$sign\_before < 0$} \Comment Follows Convexity
        \State $bound \gets y_{i} - t$
    \Else \Comment{Follows Concavity}
        \State $bound \gets y_{i} + t$  
    \EndIf
    
    \If{$Limit\,Condition$} \Comment{if condition matches Zero is applied, otherwise, Linear}
        \State $y_{i+\frac{1}{2}} \gets \frac{y_{i-1} + bound}{2}$ \Comment{Choosing middle value between Zero interpolation and the bound}
        \State $a_0 \gets \frac{y_{i+\frac{1}{2}}-y_{i-2}}{x_{i+\frac{1}{2}}-x_{i-1}}$ \Comment{Linear interpolation between $x_i$ and $x_{i+\frac{1}{2}}$}
        \State $b_0 \gets y_{i} - a_0x_{i}$
        \State $a_1 \gets \frac{y_{i-1}-y_{i+\frac{1}{2}}}{x_{i,n} - x_{i+\frac{1}{2}}} $ \Comment{We assign the value of Zero interpolation to the last point that we interpolate}
        \State $b_1 \gets y_{i+\frac{1}{2}} - a_1x_{i+\frac{1}{2}}$
    \Else
        \State $y_{i+\frac{1}{2}} \gets \frac{ax_{i+\frac{1}{2}} + b + bound}{2}$ \Comment{Choosing middle value between Linear interpolation and the bound}
        \State $a_0 \gets \frac{y_{i+\frac{1}{2}}-y_{i-2}}{x_{i+\frac{1}{2}}-x_{i-1}}$ \Comment{Linear interpolation between $x_i$ and $x_{i+\frac{1}{2}}$}
        \State $b_0 \gets y_{i} - a_0x_{i}$
        \State $a_1 \gets \frac{y_{i+1}-y_{i+\frac{1}{2}}}{x_{i+1} - x_{i+\frac{1}{2}}} $ \Comment{\textbf{Mat: needed here}}
        \State $b_1 \gets y_{i+\frac{1}{2}} - a_1x_{i+\frac{1}{2}}$
    \EndIf
            
    \State $
        \forall j=1,\dots ,n \quad y_{i,j} \gets 
        \begin{cases*}
                a_0x_{i,j}+b_0 & if $x_{i,j} \leq x_{i+\frac{1}{2}}$\\
                a_1x_{i,j}+b_1 & if $x_{i,j} > x_{i+\frac{1}{2}}$\\
        \end{cases*}
        $
\Else
    \State $a \gets \frac{y_{i+1}-y_i}{x_{i+1}-x_i}$\Comment From Linear interpolation \Comment{Apply ZeLi algorithm}
\State $Limit\,Condition \gets x_{i,n}>t\left|\frac{1}{a}\right|+x_i$ \eqref{limit_cond}
\If{Limit Condition}
\State $y_{i,j}=y_i$ for $j=0,\dots,n$ \Comment ZOH interpolation is applied
\Else
\State $b \gets y_i - ax_i$ \Comment From Linear interpolation
\State $y_{i,j}=ax_{i,j}+b$ for $j=0,\dots,n$ \Comment Linear interpolation is applied
\EndIf
\EndIf
\end{algorithmic}
\end{algorithm}

\end{document}